\documentclass[a4paper,11pt]{article}

\pdfoutput=1 

\usepackage{jinstpub} 

\usepackage{jinstpub}
\usepackage{multicol}
\usepackage{multirow}
\usepackage{orcidlink}

\usepackage{gensymb}
\usepackage{array}
\newcolumntype{x}[1]{>{\centering\arraybackslash\hspace{0pt}}p{#1}}
\usepackage[utf8]{inputenc}
\usepackage{graphicx}
\usepackage[labelformat=simple]{subcaption}

\usepackage{xcolor} 
\usepackage{soul}
\usepackage{lineno}
\usepackage{float}
\usepackage{mathrsfs}
\usepackage{amsmath}
\usepackage{array,tabularx}   
\usepackage{siunitx}
\usepackage{comment}
\def\Put(#1,#2)#3{\leavevmode\makebox(0,0){\put(#1,#2){#3}}}
\usepackage{sectsty}
\allsectionsfont{\boldmath}

\newenvironment{conditions*} 
  {\par\vspace{\abovedisplayskip}\noindent
   \tabularx{\columnwidth}{>{$}l<{$} @{${}={}$} >{\raggedright\arraybackslash}X}}
  {\endtabularx\par\vspace{\belowdisplayskip}}
\title{Spatial and Temporal Evaluations of the Liquid Argon Purity in ProtoDUNE-SP}

\collaboration{The DUNE Collaboration}
\collaboration{The DUNE Collaboration}
\affiliation[0]{University of Albany, SUNY, Albany, NY 12222, USA}
\affiliation[1]{Institute of Nuclear Physics at Almaty, Almaty 050032, Kazakhstan
}
\affiliation[2]{University of Amsterdam, NL-1098 XG Amsterdam, The Netherlands}
\affiliation[3]{Antalya Bilim University, 07190 D\"o{\c s}emealtı/Antalya, Turkey}
\affiliation[4]{University of Antananarivo, Antananarivo 101, Madagascar}
\affiliation[5]{University of Antioquia, Medell\'in, Colombia}
\affiliation[6]{Universidad Antonio Nari\~no, Bogot\'a, Colombia}
\affiliation[7]{Argonne National Laboratory, Argonne, IL 60439, USA}
\affiliation[8]{University of Arizona, Tucson, AZ 85721, USA}
\affiliation[9]{Universidad Nacional de Asunci\'on, San Lorenzo, Paraguay}
\affiliation[10]{University of Athens, Zografou GR 157 84, Greece}
\affiliation[11]{Universidad del Atl\'antico, Barranquilla, Atl\'antico, Colombia}
\affiliation[12]{Augustana University, Sioux Falls, SD 57197, USA}
\affiliation[13]{University of Bern, CH-3012 Bern, Switzerland}
\affiliation[14]{Beykent University, Istanbul, Turkey}
\affiliation[15]{University of Birmingham, Birmingham B15 2TT, United Kingdom}
\affiliation[16]{Universit\`a di Bologna, 40127 Bologna, Italy}
\affiliation[17]{Boston University, Boston, MA 02215, USA}
\affiliation[18]{University of Bristol, Bristol BS8 1TL, United Kingdom}
\affiliation[19]{Brookhaven National Laboratory, Upton, NY 11973, USA}
\affiliation[20]{University of Bucharest, Bucharest, Romania}
\affiliation[21]{University of California Berkeley, Berkeley, CA 94720, USA}
\affiliation[22]{University of California Davis, Davis, CA 95616, USA}
\affiliation[23]{University of California Irvine, Irvine, CA 92697, USA}
\affiliation[24]{University of California Los Angeles, Los Angeles, CA 90095, USA}
\affiliation[25]{University of California Riverside, Riverside CA 92521, USA}
\affiliation[26]{University of California Santa Barbara, Santa Barbara, CA 93106, USA}
\affiliation[27]{California Institute of Technology, Pasadena, CA 91125, USA}
\affiliation[28]{University of Cambridge, Cambridge CB3 0HE, United Kingdom}
\affiliation[29]{Universidade Estadual de Campinas, Campinas - SP, 13083-970, Brazil}
\affiliation[30]{Universit\`a di Catania, 2 - 95131 Catania, Italy}
\affiliation[31]{Universidad Cat\'olica del Norte, Antofagasta, Chile}
\affiliation[32]{Centro Brasileiro de Pesquisas F\'isicas, Rio de Janeiro, RJ 22290-180, Brazil}
\affiliation[33]{IRFU, CEA, Universit\'e Paris-Saclay, F-91191 Gif-sur-Yvette, France}
\affiliation[34]{CERN, The European Organization for Nuclear Research, 1211 Meyrin, Switzerland}
\affiliation[35]{Institute of Particle and Nuclear Physics of the Faculty of Mathematics and Physics of the Charles University, 180 00 Prague 8, Czech Republic }
\affiliation[36]{University of Chicago, Chicago, IL 60637, USA}
\affiliation[37]{Chung-Ang University, Seoul 06974, South Korea}
\affiliation[38]{CIEMAT, Centro de Investigaciones Energ\'eticas, Medioambientales y Tecnol\'ogicas, E-28040 Madrid, Spain}
\affiliation[39]{University of Cincinnati, Cincinnati, OH 45221, USA}
\affiliation[40]{Centro de Investigaci\'on y de Estudios Avanzados del Instituto Polit\'ecnico Nacional (Cinvestav), Mexico City, Mexico}
\affiliation[41]{Universidad de Colima, Colima, Mexico}
\affiliation[42]{University of Colorado Boulder, Boulder, CO 80309, USA}
\affiliation[43]{Colorado State University, Fort Collins, CO 80523, USA}
\affiliation[44]{Columbia University, New York, NY 10027, USA}
\affiliation[45]{Comisi\'on Nacional de Investigaci\'on y Desarrollo Aeroespacial, Lima, Peru}
\affiliation[46]{Centro de Tecnologia da Informacao Renato Archer, Amarais - Campinas, SP - CEP 13069-901}
\affiliation[47]{Central University of South Bihar, Gaya, 824236, India
}
\affiliation[48]{Institute of Physics, Czech Academy of Sciences, 182 00 Prague 8, Czech Republic}
\affiliation[49]{Czech Technical University, 115 19 Prague 1, Czech Republic}
\affiliation[50]{Laboratoire d'Annecy de Physique des Particules, Universit\'e Savoie Mont Blanc, CNRS, LAPP-IN2P3, 74000 Annecy, France}
\affiliation[51]{Daresbury Laboratory, Cheshire WA4 4AD, United Kingdom}
\affiliation[52]{Dordt University, Sioux Center, IA 51250, USA}
\affiliation[53]{Drexel University, Philadelphia, PA 19104, USA}
\affiliation[54]{Duke University, Durham, NC 27708, USA}
\affiliation[55]{University of Edinburgh, Edinburgh EH8 9YL, United Kingdom}
\affiliation[56]{Universidad EIA, Envigado, Antioquia, Colombia}
\affiliation[57]{E\"otv\"os Lor\'and University, 1053 Budapest, Hungary}
\affiliation[58]{Erciyes University, Kayseri, Turkey}
\affiliation[59]{Faculdade de Ci\^encias da Universidade de Lisboa - FCUL, 1749-016 Lisboa, Portugal}
\affiliation[60]{Universidade Federal de Alfenas, Po{\c c}os de Caldas - MG, 37715-400, Brazil}
\affiliation[61]{Universidade Federal de Goias, Goiania, GO 74690-900, Brazil}
\affiliation[62]{Universidade Federal do ABC, Santo Andr\'e - SP, 09210-580, Brazil}
\affiliation[63]{Universidade Federal do Rio de Janeiro, Rio de Janeiro - RJ, 21941-901, Brazil}
\affiliation[64]{Fermi National Accelerator Laboratory, Batavia, IL 60510, USA}
\affiliation[65]{University of Ferrara, Ferrara, Italy}
\affiliation[66]{University of Florida, Gainesville, FL 32611-8440, USA}
\affiliation[67]{Florida State University, Tallahassee, FL, 32306 USA}
\affiliation[68]{Fluminense Federal University, 9 Icara\'i Niter\'oi - RJ, 24220-900, Brazil }
\affiliation[69]{Universit\`a degli Studi di Genova, Genova, Italy}
\affiliation[70]{Georgian Technical University, Tbilisi, Georgia}
\affiliation[71]{University of Granada \& CAFPE, 18002 Granada, Spain}
\affiliation[72]{Gran Sasso Science Institute, L'Aquila, Italy}
\affiliation[73]{Laboratori Nazionali del Gran Sasso, L'Aquila AQ, Italy}
\affiliation[74]{University Grenoble Alpes, CNRS, Grenoble INP, LPSC-IN2P3, 38000 Grenoble, France}
\affiliation[75]{Universidad de Guanajuato, Guanajuato, C.P. 37000, Mexico}
\affiliation[76]{Harish-Chandra Research Institute, Jhunsi, Allahabad 211 019, India}
\affiliation[77]{University of Hawaii, Honolulu, HI 96822, USA}
\affiliation[78]{Hong Kong University of Science and Technology, Kowloon, Hong Kong, China}
\affiliation[79]{University of Houston, Houston, TX 77204, USA}
\affiliation[80]{University of  Hyderabad, Gachibowli, Hyderabad - 500 046, India}
\affiliation[81]{Idaho State University, Pocatello, ID 83209, USA}
\affiliation[82]{Instituto de F\'isica Corpuscular, CSIC and Universitat de Val\`encia, 46980 Paterna, Valencia, Spain}
\affiliation[83]{Instituto Galego de F\'isica de Altas Enerx\'ias, University of Santiago de Compostela, Santiago de Compostela, 15782, Spain}
\affiliation[84]{Institute of High Energy Physics, Chinese Academy of Sciences, Beijing, China}
\affiliation[85]{Indian Institute of Technology Kanpur, Uttar Pradesh 208016, India}
\affiliation[86]{Illinois Institute of Technology, Chicago, IL 60616, USA}
\affiliation[87]{Imperial College of Science, Technology and Medicine, London SW7 2BZ, United Kingdom}
\affiliation[88]{Indian Institute of Technology Guwahati, Guwahati, 781 039, India}
\affiliation[89]{Indian Institute of Technology Hyderabad, Hyderabad, 502285, India}
\affiliation[90]{Indiana University, Bloomington, IN 47405, USA}
\affiliation[91]{Istituto Nazionale di Fisica Nucleare Sezione di Bologna, 40127 Bologna BO, Italy}
\affiliation[92]{Istituto Nazionale di Fisica Nucleare Sezione di Catania, I-95123 Catania, Italy}
\affiliation[93]{Istituto Nazionale di Fisica Nucleare Sezione di Ferrara, I-44122 Ferrara, Italy}
\affiliation[94]{Istituto Nazionale di Fisica Nucleare Laboratori Nazionali di Frascati, Frascati, Roma, Italy}
\affiliation[95]{Istituto Nazionale di Fisica Nucleare Sezione di Genova, 16146 Genova GE, Italy}
\affiliation[96]{Istituto Nazionale di Fisica Nucleare Sezione di Lecce, 73100 - Lecce, Italy}
\affiliation[97]{Istituto Nazionale di Fisica Nucleare Sezione di Milano Bicocca, 3 - I-20126 Milano, Italy}
\affiliation[98]{Istituto Nazionale di Fisica Nucleare Sezione di Milano, 20133 Milano, Italy}
\affiliation[99]{Istituto Nazionale di Fisica Nucleare Sezione di Napoli, I-80126 Napoli, Italy}
\affiliation[100]{Istituto Nazionale di Fisica Nucleare Sezione di Padova, 35131 Padova, Italy}
\affiliation[101]{Istituto Nazionale di Fisica Nucleare Sezione di Pavia,  I-27100 Pavia, Italy}
\affiliation[102]{Istituto Nazionale di Fisica Nucleare Laboratori Nazionali di Pisa, Pisa PI, Italy}
\affiliation[103]{Istituto Nazionale di Fisica Nucleare Sezione di Roma, 00185 Roma RM, Italy}
\affiliation[104]{Istituto Nazionale di Fisica Nucleare Roma Tor Vergata , 00133 Roma RM, Italy}
\affiliation[105]{Istituto Nazionale di Fisica Nucleare Laboratori Nazionali del Sud, 95123 Catania, Italy}
\affiliation[106]{Istituto Nazionale di Fisica Nucleare, Sezione di Torino, Turin, Italy}
\affiliation[107]{Universidad Nacional de Ingenier\'ia, Lima 25, Per\'u}
\affiliation[108]{University of Insubria, Via Ravasi, 2, 21100 Varese VA, Italy}
\affiliation[109]{University of Iowa, Iowa City, IA 52242, USA}
\affiliation[110]{Iowa State University, Ames, Iowa 50011, USA}
\affiliation[111]{Institut de Physique des 2 Infinis de Lyon, 69622 Villeurbanne, France}
\affiliation[112]{Institute for Research in Fundamental Sciences, Tehran, Iran}
\affiliation[113]{Particle Physics and Cosmology International Research Laboratory	, Chicago IL,  60637 USA}
\affiliation[114]{Instituto Superior T\'ecnico - IST, Universidade de Lisboa, 1049-001 Lisboa, Portugal}
\affiliation[115]{Instituto Tecnol\'ogico de Aeron\'autica, Sao Jose dos Campos, Brazil}
\affiliation[116]{Iwate University, Morioka, Iwate 020-8551, Japan}
\affiliation[117]{Jackson State University, Jackson, MS 39217, USA}
\affiliation[118]{Jawaharlal Nehru University, New Delhi 110067, India}
\affiliation[119]{Jeonbuk National University, Jeonrabuk-do 54896, South Korea}
\affiliation[120]{Jyv\"askyl\"a University, FI-40014 Jyv\"askyl\"a, Finland}
\affiliation[121]{Kansas State University, Manhattan, KS 66506, USA}
\affiliation[122]{Kavli Institute for the Physics and Mathematics of the Universe, Kashiwa, Chiba 277-8583, Japan}
\affiliation[123]{High Energy Accelerator Research Organization (KEK), Ibaraki, 305-0801, Japan}
\affiliation[124]{Korea Institute of Science and Technology Information, Daejeon, 34141, South Korea}
\affiliation[125]{Taras Shevchenko National University of Kyiv, 01601 Kyiv, Ukraine}
\affiliation[126]{Lancaster University, Lancaster LA1 4YB, United Kingdom}
\affiliation[127]{Lawrence Berkeley National Laboratory, Berkeley, CA 94720, USA}
\affiliation[128]{Laborat\'orio de Instrumenta{\c c}\~ao e F\'isica Experimental de Part\'iculas, 1649-003 Lisboa and 3004-516 Coimbra, Portugal}
\affiliation[129]{University of Liverpool, L69 7ZE, Liverpool, United Kingdom}
\affiliation[130]{Los Alamos National Laboratory, Los Alamos, NM 87545, USA}
\affiliation[131]{Louisiana State University, Baton Rouge, LA 70803, USA}
\affiliation[132]{Laboratoire de Physique des Deux Infinis Bordeaux - IN2P3, F-33175 Gradignan, Bordeaux, France, }
\affiliation[133]{University of Lucknow, Uttar Pradesh 226007, India}
\affiliation[134]{Johannes Gutenberg-Universit\"at Mainz, 55122 Mainz, Germany}
\affiliation[135]{University of Manchester, Manchester M13 9PL, United Kingdom}
\affiliation[136]{Massachusetts Institute of Technology, Cambridge, MA 02139, USA}
\affiliation[137]{University of Medell\'in, Medell\'in, 050026 Colombia }
\affiliation[138]{University of Michigan, Ann Arbor, MI 48109, USA}
\affiliation[139]{Michigan State University, East Lansing, MI 48824, USA}
\affiliation[140]{Universit\`a di Milano Bicocca , 20126 Milano, Italy}
\affiliation[141]{Universit\`a degli Studi di Milano, I-20133 Milano, Italy}
\affiliation[142]{University of Minnesota Duluth, Duluth, MN 55812, USA}
\affiliation[143]{University of Minnesota Twin Cities, Minneapolis, MN 55455, USA}
\affiliation[144]{University of Mississippi, University, MS 38677 USA}
\affiliation[145]{Universit\`a degli Studi di Napoli Federico II , 80138 Napoli NA, Italy}
\affiliation[146]{Nikhef National Institute of Subatomic Physics, 1098 XG Amsterdam, Netherlands}
\affiliation[147]{National Institute of Science Education and Research (NISER), Odisha 752050, India}
\affiliation[148]{University of North Dakota, Grand Forks, ND 58202-8357, USA}
\affiliation[149]{Northern Illinois University, DeKalb, IL 60115, USA}
\affiliation[150]{Northwestern University, Evanston, Il 60208, USA}
\affiliation[151]{University of Notre Dame, Notre Dame, IN 46556, USA}
\affiliation[152]{University of Novi Sad, 21102 Novi Sad, Serbia}
\affiliation[153]{Ohio State University, Columbus, OH 43210, USA}
\affiliation[154]{Oregon State University, Corvallis, OR 97331, USA}
\affiliation[155]{University of Oxford, Oxford, OX1 3RH, United Kingdom}
\affiliation[156]{Pacific Northwest National Laboratory, Richland, WA 99352, USA}
\affiliation[157]{Universt\`a degli Studi di Padova, I-35131 Padova, Italy}
\affiliation[158]{Panjab University, Chandigarh, 160014, India}
\affiliation[159]{Universit\'e Paris-Saclay, CNRS/IN2P3, IJCLab, 91405 Orsay, France}
\affiliation[160]{Universit\'e Paris Cit\'e, CNRS, Astroparticule et Cosmologie, Paris, France}
\affiliation[161]{University of Parma,  43121 Parma PR, Italy}
\affiliation[162]{Universit\`a degli Studi di Pavia, 27100 Pavia PV, Italy}
\affiliation[163]{University of Pennsylvania, Philadelphia, PA 19104, USA}
\affiliation[164]{Pennsylvania State University, University Park, PA 16802, USA}
\affiliation[165]{Physical Research Laboratory, Ahmedabad 380 009, India}
\affiliation[166]{Universit\`a di Pisa, I-56127 Pisa, Italy}
\affiliation[167]{University of Pittsburgh, Pittsburgh, PA 15260, USA}
\affiliation[168]{Pontificia Universidad Cat\'olica del Per\'u, Lima, Per\'u}
\affiliation[169]{University of Puerto Rico, Mayaguez 00681, Puerto Rico, USA}
\affiliation[170]{Punjab Agricultural University, Ludhiana 141004, India}
\affiliation[171]{Queen Mary University of London, London E1 4NS, United Kingdom
}
\affiliation[172]{Radboud University, NL-6525 AJ Nijmegen, Netherlands}
\affiliation[173]{Rice University, Houston, TX 77005}
\affiliation[174]{University of Rochester, Rochester, NY 14627, USA}
\affiliation[175]{Royal Holloway College London, London, TW20 0EX, United Kingdom}
\affiliation[176]{Rutgers University, Piscataway, NJ, 08854, USA}
\affiliation[177]{STFC Rutherford Appleton Laboratory, Didcot OX11 0QX, United Kingdom}
\affiliation[178]{Universit\`a del Salento, 73100 Lecce, Italy}
\affiliation[179]{Universidad del Magdalena, Santa Marta - Colombia}
\affiliation[180]{Sapienza University of Rome, 00185 Roma RM, Italy}
\affiliation[181]{Universidad Sergio Arboleda, 11022 Bogot\'a, Colombia}
\affiliation[182]{University of Sheffield, Sheffield S3 7RH, United Kingdom}
\affiliation[183]{SLAC National Accelerator Laboratory, Menlo Park, CA 94025, USA}
\affiliation[184]{University of South Carolina, Columbia, SC 29208, USA}
\affiliation[185]{South Dakota School of Mines and Technology, Rapid City, SD 57701, USA}
\affiliation[186]{South Dakota State University, Brookings, SD 57007, USA}
\affiliation[187]{Stony Brook University, SUNY, Stony Brook, NY 11794, USA}
\affiliation[188]{Sanford Underground Research Facility, Lead, SD, 57754, USA}
\affiliation[189]{University of Sussex, Brighton, BN1 9RH, United Kingdom}
\affiliation[190]{Syracuse University, Syracuse, NY 13244, USA}
\affiliation[191]{Universidade Tecnol\'ogica Federal do Paran\'a, Curitiba, Brazil}
\affiliation[192]{Tel Aviv University, Tel Aviv-Yafo, Israel}
\affiliation[193]{Texas A\&M University, College Station, Texas 77840}
\affiliation[194]{Texas A\&M University - Corpus Christi, Corpus Christi, TX 78412, USA}
\affiliation[195]{University of Texas at Arlington, Arlington, TX 76019, USA}
\affiliation[196]{University of Texas at Austin, Austin, TX 78712, USA}
\affiliation[197]{University of Toronto, Toronto, Ontario M5S 1A1, Canada}
\affiliation[198]{Tufts University, Medford, MA 02155, USA}
\affiliation[199]{Universidade Federal de S\~ao Paulo, 09913-030, S\~ao Paulo, Brazil}
\affiliation[200]{Ulsan National Institute of Science and Technology, Ulsan 689-798, South Korea}
\affiliation[201]{University College London, London, WC1E 6BT, United Kingdom}
\affiliation[202]{University of Kansas, Lawrence, KS 66045}
\affiliation[203]{Universidad Nacional Mayor de San Marcos, Lima, Peru}
\affiliation[204]{Valley City State University, Valley City, ND 58072, USA}
\affiliation[205]{University of Vigo, E- 36310 Vigo Spain}
\affiliation[206]{Virginia Tech, Blacksburg, VA 24060, USA}
\affiliation[207]{University of Warsaw, 02-093 Warsaw, Poland}
\affiliation[208]{University of Warwick, Coventry CV4 7AL, United Kingdom}
\affiliation[209]{Wellesley College, Wellesley, MA 02481, USA}
\affiliation[210]{Wichita State University, Wichita, KS 67260, USA}
\affiliation[211]{William and Mary, Williamsburg, VA 23187, USA}
\affiliation[212]{University of Wisconsin Madison, Madison, WI 53706, USA}
\affiliation[213]{Yale University, New Haven, CT 06520, USA}
\affiliation[214]{Yerevan Institute for Theoretical Physics and Modeling, Yerevan 0036, Armenia}
\affiliation[215]{York University, Toronto M3J 1P3, Canada}
\author[112]{S.~Abbaslu,}
\author[34]{A.~Abed Abud,}
\author[64]{R.~Acciarri,}
\author[191]{L.~P.~Accorsi,}
\author[11]{M.~A.~Acero,}
\author[191]{M.~R.~Adames,}
\author[70]{G.~Adamov,}
\author[64]{M.~Adamowski,}
\author[29]{C.~Adriano,}
\author[174]{F.~Akbar,}
% ...

\emailAdd{bianjm@uci.edu}

\abstract{Liquid argon time projection chambers (LArTPCs) rely on highly pure argon to ensure that ionization electrons produced by charged particles reach readout arrays. ProtoDUNE Single-Phase (ProtoDUNE-SP) was an approximately 700-ton liquid argon detector intended to prototype the Deep Underground Neutrino Experiment (DUNE) Far Detector Horizontal Drift module. It contains two drift volumes bisected by the cathode plane assembly, which is biased to create an almost uniform electric field in both volumes. The DUNE Far Detector modules must have robust cryogenic systems capable of filtering argon and supplying the TPC with clean liquid. This paper will explore comparisons of the argon purity measured by the purity monitors with those measured using muons in the TPC from October 2018 to November 2018. A new method is introduced to measure the liquid argon purity in the TPC using muons crossing both drift volumes of ProtoDUNE-SP. For extended periods on the timescale of weeks, the drift electron lifetime was measured to be above 30 ms using both systems. A particular focus will be placed on the measured purity of argon as a function of position in the detector.
}


\arxivnumber{2507.08586} 

\begin{document}
\hfill FERMILAB-PUB-25-0445-V

\hfill CERN-EP-2025-157

\maketitle
\flushbottom

\section{Introduction} \label{sec:intro}
A liquid argon time projection chamber (LArTPC) detects charged particles traveling through liquid argon by collecting ionization electrons that drift toward readout sensors~\cite{Nygren:1974nfi,Rubbia:1977zz}. The technology is ideal for MeV-scale physics, due to liquid argon's high yield of approximately 42 ke$^-$/MeV~\cite{Miyajima:1974zz}. The ionization electrons sometimes drift through several meters of argon to reach the sensors, as expected for signals in the proposed Deep Underground Neutrino Experiment (DUNE) Far Detector Horizontal Drift module with its 3.5 m-long drift distance~\cite{duneTDRVol4}. 

ProtoDUNE Single-Phase (ProtoDUNE-SP) was a full-scale engineering prototype of the DUNE Far Detector Horizontal Drift module. ProtoDUNE-SP operated at the CERN Neutrino Platform from 2018 to 2020. While ProtoDUNE-SP had argon recirculation and filtration systems, electronegative impurities may still be present in the argon~\cite{protoDUNEDesign}. These electronegative impurities, such as oxygen and water, can capture the ionization electrons and attenuate the signals for a LArTPC. The loss of charge per unit time due to impurities ($\frac{dQ}{dt}$) is:  

\begin{equation}
    {\frac{dQ}{dt}=-k_{\rm{a}}(\mathscr{E})\rho Q},
    \label{eqn:attn}
\end{equation}

\noindent where $\rho$ is the concentration of an impurity, $k_{\rm{a}}$ is the attachment rate, which is dependent on the electric field ($\mathscr{E}$), and $Q$ is the instantaneous ionization charge~\cite{bakale, LArReview, thorn}.

Quantifying this effect allows the loss of charge due to impurities to be measured and accounted for in calibrations. This effect is typically measured as a function of drift distance and is represented as a drift electron lifetime ($\tau$=1/[$\rm{k_a \rho}$])~\cite{bakale, LArReview, instruments5010002}. The integration of Equation~\ref{eqn:attn}, assuming $k_{\rm{a}}$ and $\rho$ are constants, results in an exponential dependence of the surviving charge on the drift time: 

\begin{equation}
    Q(t_{\rm{drift}})=Q_{0}\exp(-t_{\rm{drift}}/\tau),
    \label{eqn:lifetime}
\end{equation}
where ${t_{\rm{drift}}}$ is the drift time and ${Q_0}$ is the initial ionization charge.

A low drift electron lifetime corresponds to a significant amount of charge being lost while drifting to the readout sensors. In those operating conditions, the ionization electrons may not reach the readout wires, leading to an inability to detect MeV-scale energy deposits from neutrino-induced charged particles. The DUNE Far Detector Horizontal Drift module has set technical specifications for liquid argon purity to prevent such a scenario~\cite{duneTDRVol2}. These benchmarks depend on the maximum drift time of the detector and the inferred primary contaminant of the TPC, which for DUNE is oxygen. ProtoDUNE-SP has a maximum drift time from cathode to anode of approximately 2.25 ms. The DUNE Far Detector Horizontal Drift module is expected to have a similar maximum drift time. Table~\ref{tab:Requirements} shows the related argon purity requirements and specifications. A more common quantity for evaluating the purity in a LArTPC is the equivalent oxygen (O$_2$) concentration. The concentration is acquired from the drift electron lifetime and oxygen equivalent attachment rate, which is $\rm{\frac{1}{300 \, ppt \, ms}}$ at the DUNE Far Detector's specified electric field of 500 V/cm. The attachment rate is determined from previous studies by Bakale~\cite{bakale} and ICARUS \cite{bettini}, as parameterized in~\cite{thorn}. 

\begin{table}[H]
    \centering
        \caption{DUNE Far Detector Horizontal Drift module technical benchmarks for liquid argon purity quantified in various ways~\cite{duneTDRVol2}.}
    \begin{tabular}{c|c|c}
      Metric   & Requirement & Specification  \\ \hline
      Drift electron lifetime &  3 ms & 10 ms  \\ \hline
     Maximum drift charge lost    & 54\% & 21\% \\ \hline
     $\rm{O_2}$ equivalent concentration & 100 ppt & 30 ppt \\ \hline 
    \end{tabular}
    \label{tab:Requirements}
\end{table}

Previous measurements of the drift electron lifetime in the TPC using muons yielded lifetimes exceeding 20 ms during long periods of ProtoDUNE-SP operation~\cite{protoDUNEDesign,abi2020first}. These results were confirmed by measurements from purity monitors adapted from the same designs employed by ICARUS~\cite{icarus}. The purity monitor data reported electron lifetime consistently greater than 30 ms over a nearly two-year period~\cite{protoDUNEDesign}.

This work intends to compare and contrast methods of measuring the drift electron lifetime using ProtoDUNE-SP data from October to November 2018. This period was chosen as it represents the longest continuous period with uniform operating conditions. This paper presents an updated analysis of the drift electron lifetime measurement using CRT-TPC matched tracks with new calibrations. It also introduces a new method for ProtoDUNE-SP to measure the drift electron lifetime using cosmic-ray muons that cross the Cathode Plane Assembly (CPA), which is the structure that provides the high voltage necessary to generate the electric field in the LArTPC. The sample of through-going cosmic-ray muons that cross the cathode provides nearly complete coverage of the detector, enabling comparisons of the liquid argon purity as a function of time and location in the detector volume. In this paper a sample of cosmic-ray muons collected by ProtoDUNE-SP is used for the first time to map the liquid argon purity as a function of the position within the TPC.

Due to the low cosmic-ray muon flux at 1.5 km underground, the statistics to cover a Far Detector module will be limited. Furthermore, the cosmic-ray muons that do reach the Far Detector module will travel vertically at shallow angles, providing good vertical coverage of the detector height but limited coverage of the drift distance and detector length. Therefore, the purity monitors will be the primary method to measure the liquid argon purity for the DUNE Far Detector modules. This work intends to confirm that the purity monitor is suitable for monitoring the liquid argon purities in the DUNE Far Detector modules. Without such a verification, the purity monitor data may lead to biases in the calorimetry and an increase in the calorimetric systematic uncertainties. Demonstrating that purity monitors yield liquid argon purity measurements consistent with those from the TPC builds confidence in their use for the DUNE Far Detector.

Section~\ref{sec:pdsp} introduces the ProtoDUNE-SP detector. Section~\ref{sec:method} discusses further how ProtoDUNE-SP measures the drift electron lifetime using the TPC, with Section~\ref{sec:PrMon} explaining the analysis with the purity monitors. Section~\ref{sec:compare} shows comparisons of the results over the relevant timescale and across the detector volume.

\section{ProtoDUNE-SP Detector} \label{sec:pdsp}

\begin{figure}
    \centering
    \includegraphics[width=0.7\textwidth]{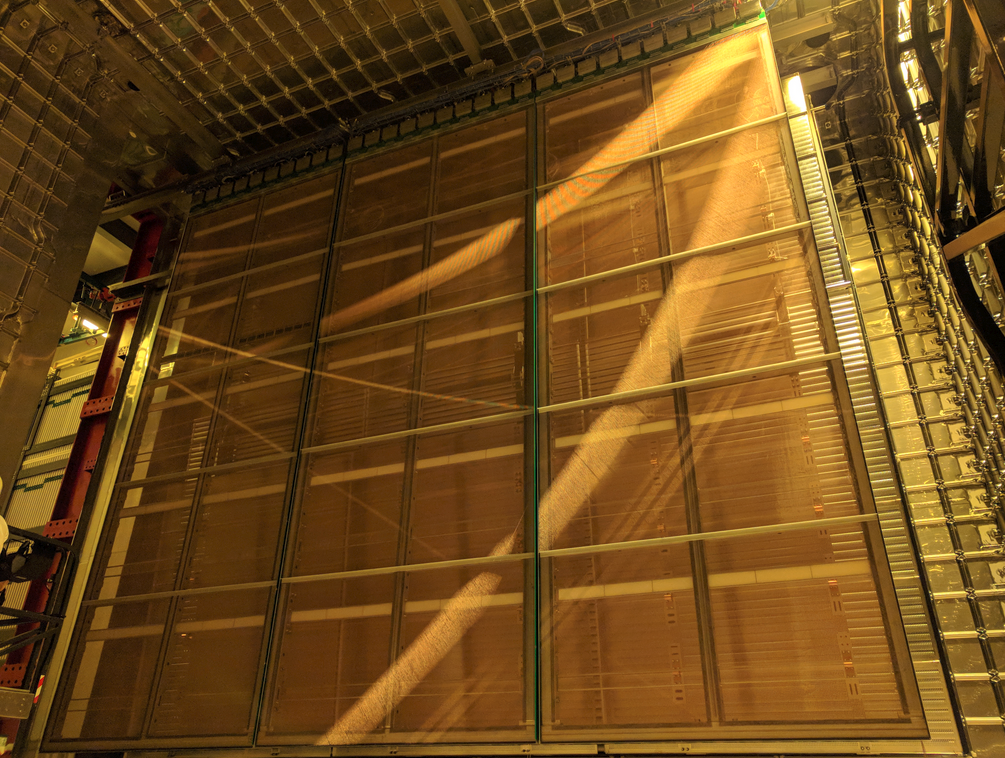}
    \caption{Three APAs in front of the field cage and cryostat wall. The image is originally from~\cite{protoDUNEDesign}.}
    \label{fig:3apas}
\end{figure}

\begin{figure}
    \centering
    \includegraphics[width=0.7\textwidth]{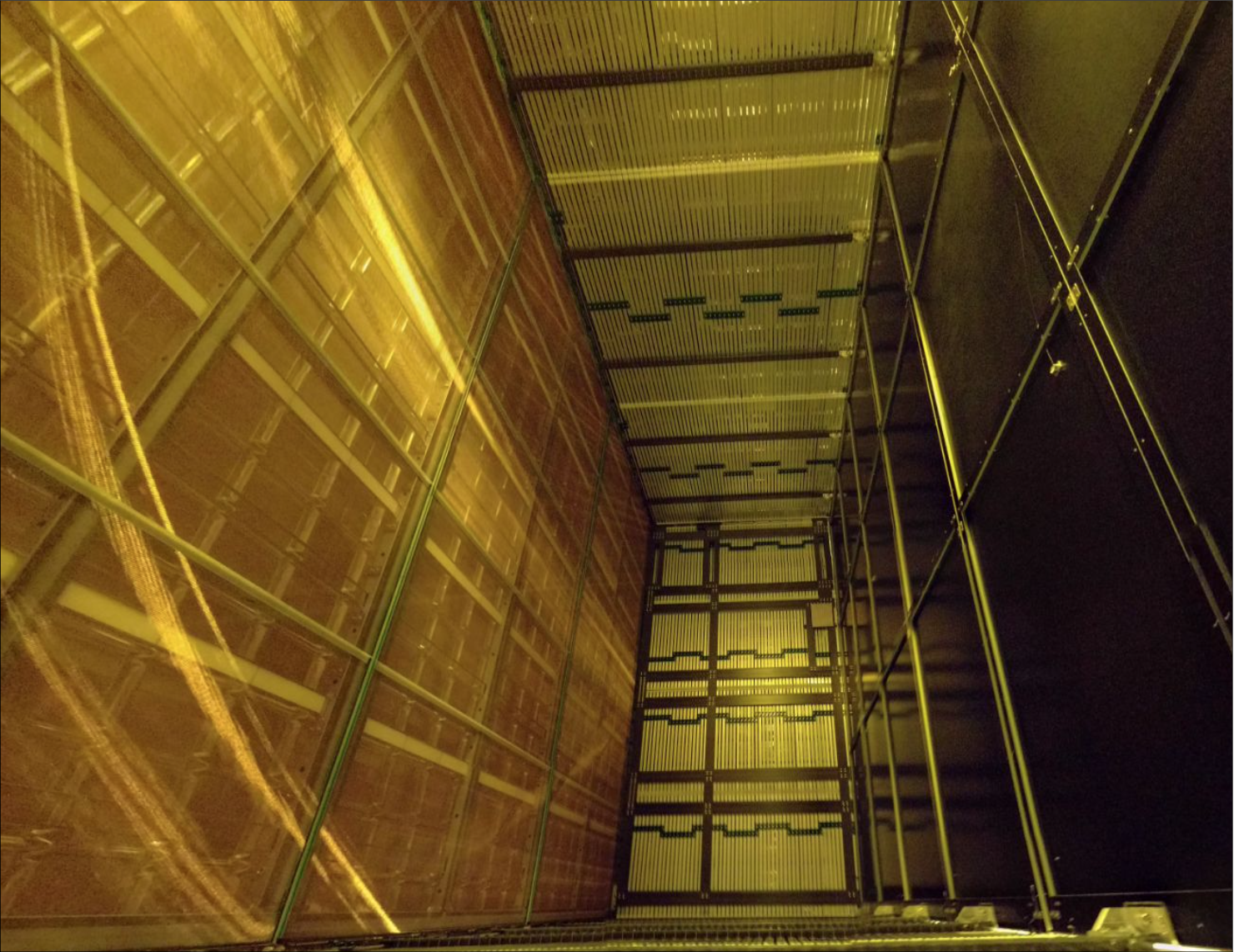}
    \caption{One drift volume with the anode (left) and cathode (right), which are 3.5 m apart. The image is originally from~\cite{abi2020first}.}
    \label{fig:drift}
\end{figure}

 The ProtoDUNE-SP detector was a LArTPC of dimensions 7.2 m wide, 6 m high, and 7 m long inside the field cage. It contains two drift volumes. Each has a 3.5 m-long drift distance and three wire-based anode plane assemblies (APAs). An image of the three APAs installed at one end of the field cage is shown in Figure~\ref{fig:3apas}. An entire drift volume is displayed in Figure~\ref{fig:drift}. An APA contains two induction planes and a collection plane. The induction planes measure ionization electrons as they drift towards and past the wires. The collection plane measures ionization electrons that gather on the wires, producing a unipolar signal on the wire. Because induction plane wires are at an angle relative to the vertical direction, only signals from the collection plane will be used for simplicity in muon selection as a function of angle.  

\begin{figure}[h]
    \centering
    \includegraphics[width=0.8\textwidth]{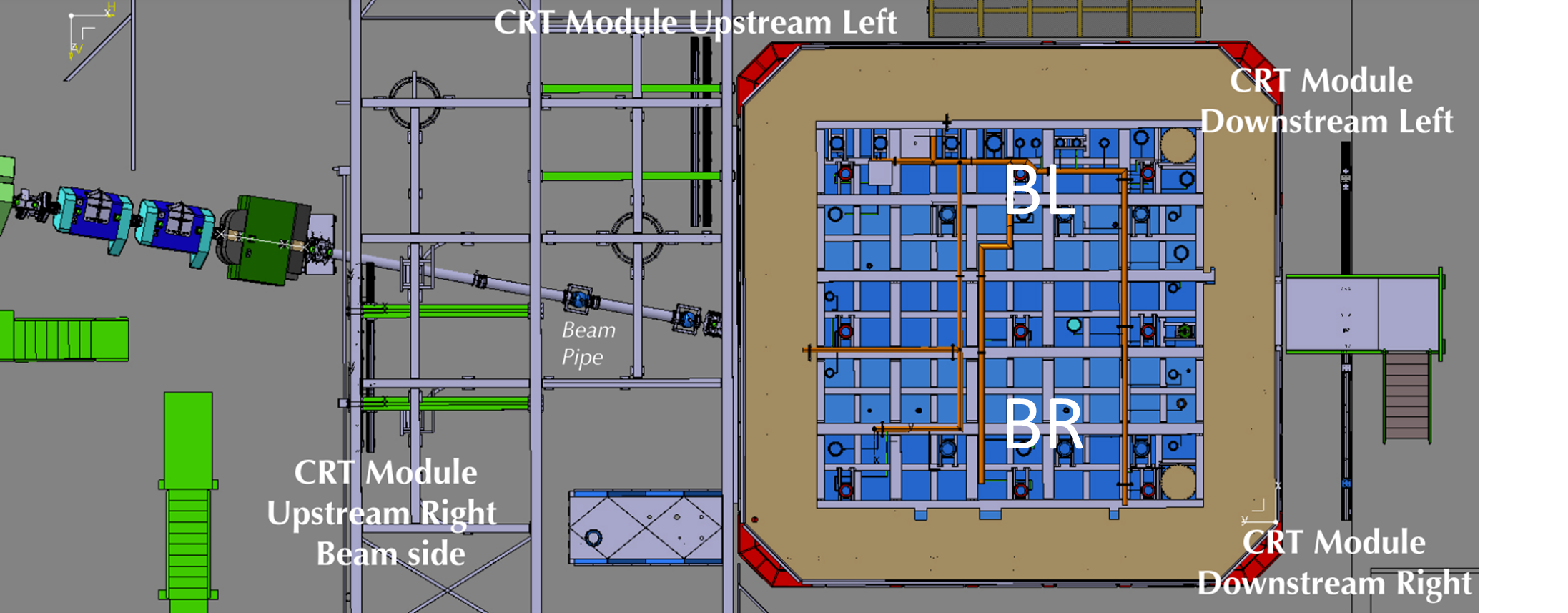}
    \caption{A diagram of the CERN Neutrino Platform area that hosted the ProtoDUNE-SP detector. The beamline instrumentation (left) and the beam pipe to the ProtoDUNE-SP cryostat (center) are included. The CRT panels cover the upstream  and downstream faces with a slight offset upstream of the detector to allow the beam pipe to enter. Beam-left (BL) is at the top of the image, and beam-right (BR) is at the bottom of this image. The diagram is from~\cite{abi2020first}.}
    \label{fig:np04}
\end{figure}

The ProtoDUNE-SP detector was exposed to the CERN H4-beamline, which provided muons, positrons, kaons, pions, and protons for measurements in liquid argon. In addition to the beam and TPC instrumentation, the upstream and downstream of ProtoDUNE-SP were surrounded by a CRT system composed of scintillator strips for muon reconstruction. The CRT provides timing information, an entry point, and an exit point that can form a vector and timestamp that can be matched to reconstructed objects in the TPC. Figure~\ref{fig:np04} shows a diagram of the ProtoDUNE-SP detector hall with the beam particles entering from left to right through the beam pipe towards the ProtoDUNE-SP cryostat~\cite{abi2020first}. The beam particles enter the TPC volume close to the CPA and almost parallel to it. Therefore, to differentiate the two drift volumes, they will be referred to as either the beam-left (BL) or the beam-right (BR) drift volumes. ProtoDUNE-SP has a right-handed coordinate system with the BL on the negative-side of the $X$ coordinate and BR on the positive-side of the $X$ coordinate. The $Z$ coordinate begins at the upstream face of the detector relative to the beam and the $Y$ coordinate is defined from floor to ceiling.

\begin{figure}[h]
        \centering
    \includegraphics[width=0.8\textwidth]{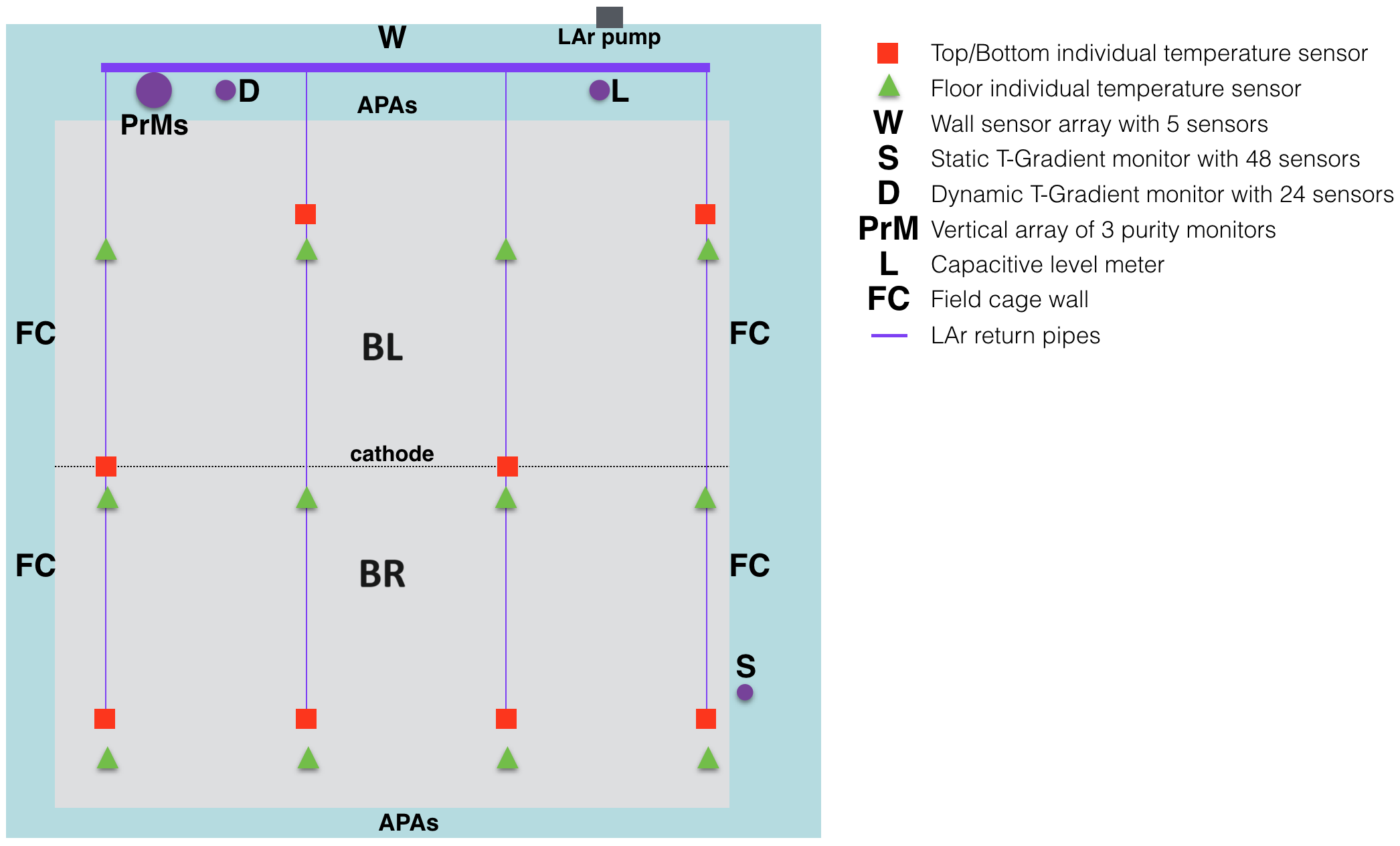}
    \caption{Top view display of the ProtoDUNE-SP field cage that highlights the location of the purity monitors relative to the detector volume~\cite{protoDUNEDesign}.}
    \label{fig:placement}
\end{figure}

\begin{figure}[h]
    \centering    \includegraphics[width=\textwidth]{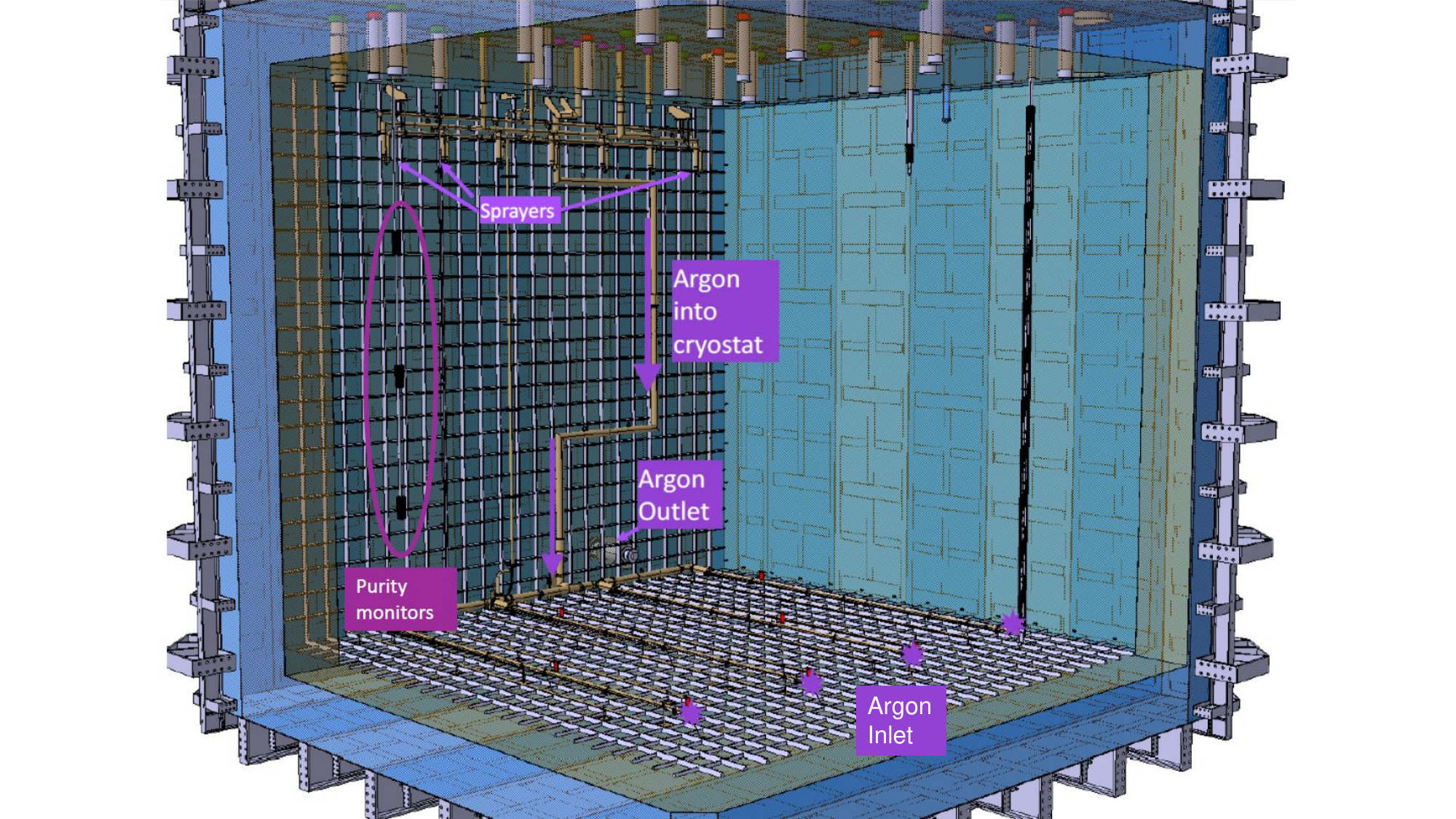}
    \caption{Diagram of the argon circulation system (CPA not shown). The sprayers are used only during filling. The purity monitors are on the left side of this diagram. The image is from~\cite{protoDUNEDesign}.}
    \label{fig:circulation}
\end{figure}

 The detector included three purity monitors outside the field cage to monitor the liquid argon purity, based on the design of ICARUS purity monitors~\cite{icarus}. A broader discussion of the operating principle of the ProtoDUNE-SP purity monitors is in Section~\ref{sec:PrMon}. Each purity monitor is installed at different heights, approximately 2 m apart, in the corner of the cryostat, as seen in Figure~\ref{fig:placement}. The placement enables redundancy in measurements and allows the purity monitors to probe the stratification of argon purity as a function of detector height. The argon outlet is positioned on one of the sidewalls to the left side of the test beam. The argon flows through the filtration and recirculation system, and purified argon returns via a network of pipes at the bottom of the cryostat~\cite{protoDUNEDesign}. Figure~\ref{fig:circulation} shows the placement of  different components of the recirculation system inside the cryostat with the cathode (not shown) in the middle of the image. Given that argon return to the cryostat is underneath the BR drift volume, it should have higher argon purity than BL. Details on the filtration and recirculation system outside the cryostat can be found in~\cite{protoDUNEDesign,abi2020first}.

\section{Methodologies for Measuring the Drift Electron Lifetime in a TPC}\label{sec:method}

The evaluation of the drift electron lifetime can be made by comparing the difference in drifting charge detected at two spatially separated points or from quantifying the attenuation of signals over a series of measurements across a drift distance. This section will describe the methods for measuring the drift electron lifetime using the TPC. The measurements use muons that traverse the detector, selected using two different event topologies. These through-going muons, with energies between 1 and 6 GeV, are sufficiently energetic to behave approximately as minimally ionizing particles (MIPs), depositing energy at an average rate of about 2 MeV/cm.. The consistent MIP deposits provide a standard candle to probe the change in ionization electrons measured as a function of drift distance and extract the drift electron lifetime~\cite{pdg}. 

\subsection{TPC Signal Reconstruction}

A several millisecond time series of analog-to-digital (ADC) samples is recorded for each readout channel (wire) on every event trigger. These waveforms are then processed using a region-of-interest finder, which finds ``hits'', normally due to the ionization charge-induced signals on the wires. The region-of-interest finder computes charge of the hit, and associates a timestamp to the hit relative to the event trigger. These hits are assembled, and contain calorimetric information, such as the charge collected per step ($dQ/dx$). The $dQ/dx$ is defined using the hit charge ($Q$) and the step length ($dx$) from one hit to the next in a reconstructed track. The wire pitch is 4.79 mm on the collection plane and $dx$ is corrected for the angle of the track relative to the wire pitch.

The hits are grouped into clusters from which the reconstruction then attempts to build the hierarchy of particles that interacted within the TPC~\cite{abi2020first,pandoraProtoDUNE}. ProtoDUNE-SP uses the Pandora reconstruction package to combine signals in the detector. Pandora is a multi-algorithmic package that uses various algorithms to provide fully reconstructed particle hierarchies representing each interaction in the detector. A detailed description of the Pandora event reconstruction applied to ProtoDUNE-SP can be found in~\cite{pandoraProtoDUNE}. 


Because the detector is exposed to a significant cosmic-ray muon flux, the local electric field is distorted from the accumulation of argon ions induced by cosmic-ray muons, affecting the reconstruction of particle trajectories and related calorimetric measurements. This is referred to as the space charge effect (SCE). It produces an electric field distortion of approximately -10\% near the anode and 20\% near the cathode for both drift volumes. A positional and electric field correction map for the SCE has been developed using through-going cosmic-ray muons~\cite{abi2020first}. The distortions of the reconstructed trajectories in the TPC are evaluated with respect to their entrance and exit points into the TPC. From these offsets, the electric field corrections are derived to form a data-driven SCE correction for calibrating the tracking and calorimetry information.

\subsection{Event Selection for CRT-TPC Matched Tracks}\label{sec:crtMethod}

Previously reported measurements~\cite{abi2020first, DiurbaThesis} of the drift electron lifetime in ProtoDUNE-SP relied on the CRTs to select particles that traverse the entire cryostat volume and have a matching track reconstructed within the TPC. Here we will show more data taken from November 2018 with additional corrections applied, present the purity measurements obtained from a new sample of cathode-crossing tracks, and provide a comparison to the results from the purity monitors. The CRT consists of 32 modules with 64 scintillator strips in each module refurbished from the Double Chooz experiment's Outer Veto~\cite{protoDUNEDesign,DoubleChooz}. Although the strips are 5 cm wide, they are assembled in two layers shifted with respect to each other to give an effective width of 2.5 cm. The modules are overlaid on one another to enable two-dimensional positional reconstruction of CRT signals using the timing information and the specific strip that triggered~\cite{protoDUNEDesign}. They have a timing resolution of 20 ns. Relative to the detector length ($Z$), half of the modules are upstream of the TPC and the other half are downstream. Drawing a line from reconstructed hits in the upstream CRT modules to the reconstructed hits in the downstream CRT modules allows for the reconstruction of candidate CRT tracks, defined as any combination of upstream hits and downstream hits within a coincidence window of 160 ns. The size of the coincidence window was chosen based on the time it takes for a relativistic muon to travel a maximum of 20 m from upstream BR CRT modules to downstream CRT modules as seen in Figure~\ref{fig:np04}, and the timing offsets between modules. The upstream modules and downstream modules are between 10-20 m apart, depending on which upstream CRT modules have signals since some are offset to avoid interfering with the placement of the ProtoDUNE-SP beam pipe (Figure~\ref{fig:np04})~\cite{abi2020first}. The CRT modules cover 6.8 m$^2$ and 6.8 m$^2$ at both upstream and downstream locations.

The CRT candidate tracks can be a calibration tool for tracking in the TPC by providing a timestamp and tracking information when matched to a suitable muon reconstructed in the TPC without needing the positional SCE calibration map. These will be referred to as CRT-TPC matched tracks. These tracks are selected by comparing the tracking and angular displacements between reconstructed muons in the TPC and the associated CRT signals~\cite{DiurbaThesis}. The TPC track and CRT track must have an angular agreement equivalent to a cosine of the angular displacement of 0.99~\cite{DiurbaThesis}. The angular displacement is calculated by taking the dot product of the CRT track direction and TPC track direction using the endpoints of each track to define their respective direction. The spatial displacement is measured by extrapolating the TPC track onto the upstream and downstream CRT modules. The displacement between the two tracks must be within 40 cm in the coordinate relative to the detector width ($X$) and 40 cm in the coordinate parallel to the detector height ($Y$). As this analysis intends to study ionization electrons produced from muons, the selection must also ensure that the muon signals do not experience any detector effects due to the angle relative to the collection wire orientation. Therefore, an additional selection step is applied requiring that the cos($\theta_{Z}$) of the track trajectory relative to the Z axis (along the detector length) must be greater than 0.99, which ensures that selected trajectories run nearly parallel to the APAs and perpendicular to the collection wires. This was done to ensure that the track pitch matched closes with the pitch of the wires to approximately 1\%. These track topologies give the most accurate calorimetric estimation of charge deposition per wire ($dQ/dx$). The CRT timestamp allows to correct the drift time coordinate for any timing offsets between the time of the event trigger and the actual entry of the muon into the TPC. The selected muons can come from either the cosmic-ray flux or the test-beam muon halo, with the only requirement that the muon travels the full distance from the upstream CRT through the TPC and the downstream CRT.

The performance of the track matching algorithm has a purity of associating the correct timestamp of greater than 99.9\%, according to simulation. The maximum angular displacement between tracks in data and simulation is no greater than 0.9996 (1.62 degrees)~\cite{DiurbaThesis}. Furthermore, the distribution of the spatial displacement ($\Delta \rm{R}$) in $X$ and $Y$ between each TPC track spacepoint and the CRT track has an approximate half-width of 10 cm~\cite{DiurbaThesis}. These metrics ensure that the CRT-TPC matched tracks agree well, despite not using positional SCE calibrations, and can be used as a well-understood sample of tracks for detector physics studies. 

\begin{figure}
    \centering
    \includegraphics[width=0.45\textwidth]{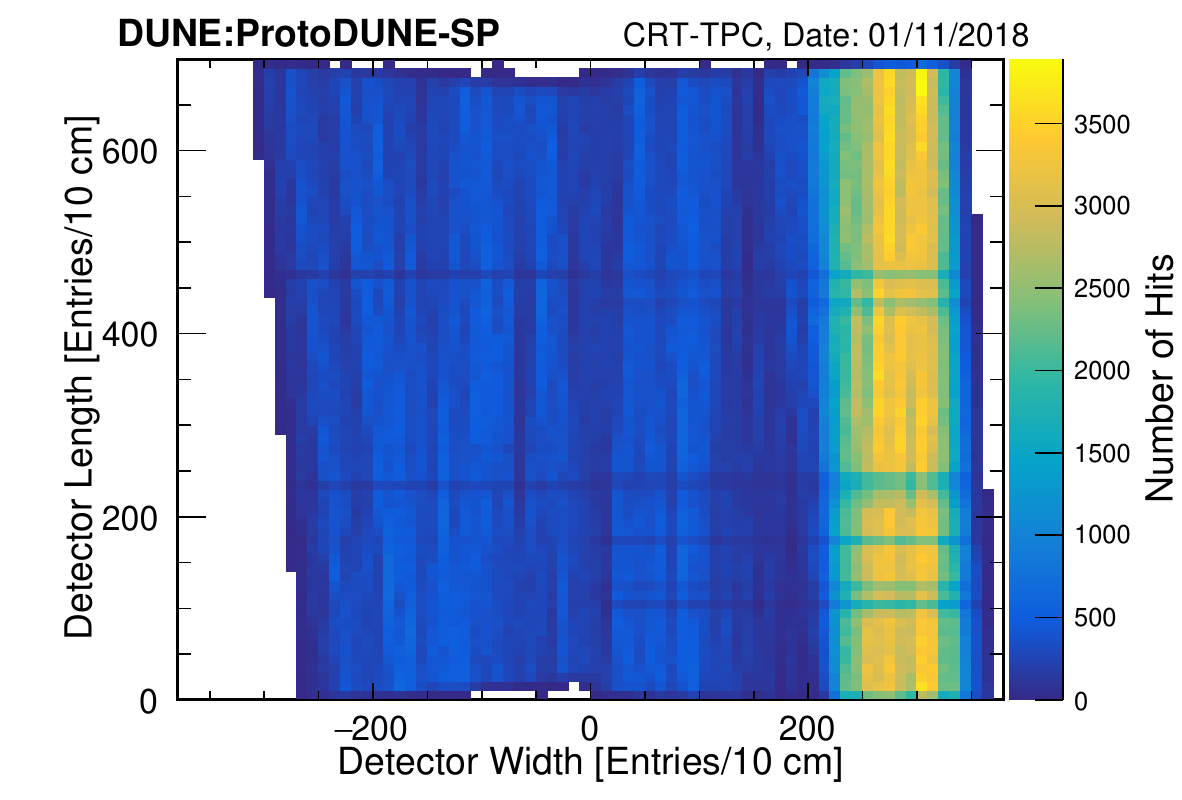}
        \includegraphics[width=0.45\textwidth]{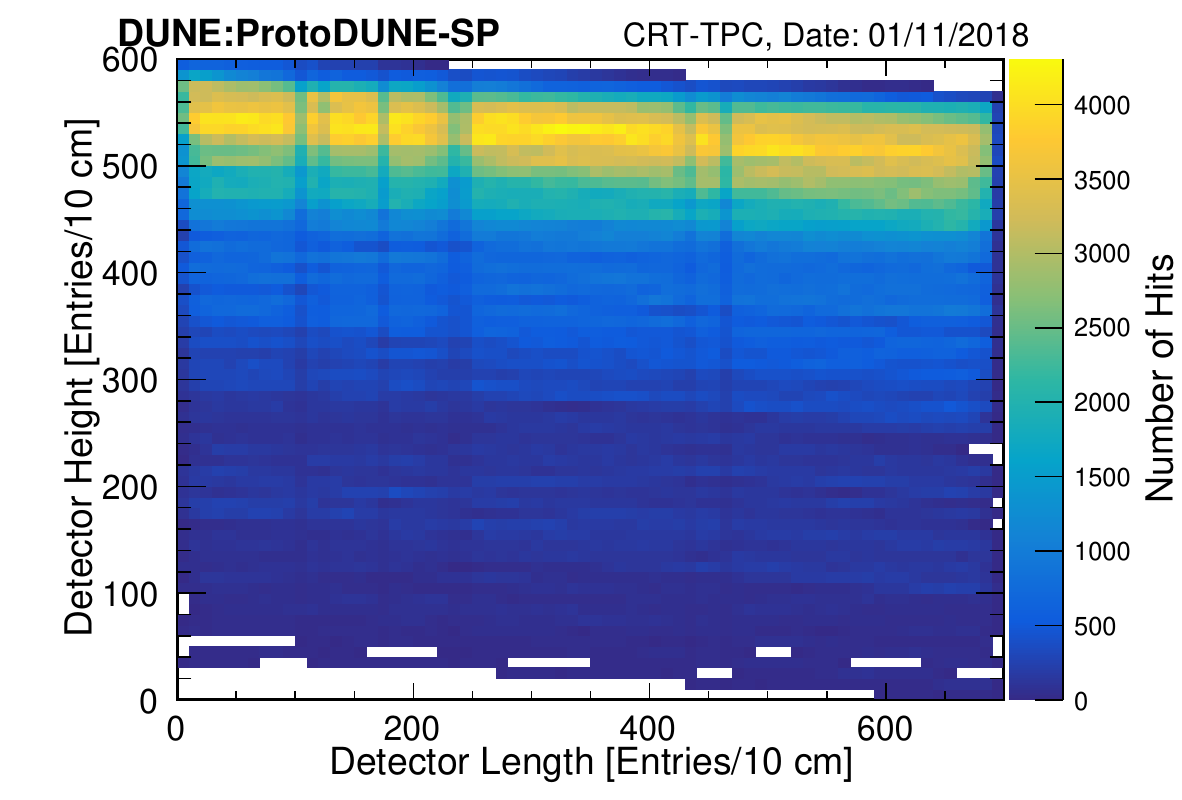}
    \caption{Statistics of hits on the collection plane from the CRT-TPC matching event selection for data taken on November 1st, 2018, from the top (left) and side view (right). The cathode is at position zero on the detector width axis, with BR at $X$ less than zero and BL at $X$ greater than zero. The position of hits are calibrated using the CRT information.}
    \label{fig:crtcoverage5758}
\end{figure}

However, due to the significant distance required for a muon to pass the upstream CRT modules, TPC, and downstream CRT modules, the statistics for these tracks are low and restricted to only certain areas of the detector, as can be seen in the coverage map in Figure~\ref{fig:crtcoverage5758}. Note that BL is actually for positions along the detector width greater than zero to ensure that a right-handed coordinate system is used. More tracks are reconstructed on the BL side since the CRT modules are closer to the TPC and exposed to beam halo muons. Complete coverage of the top and bottom of the detector is also limited since the muons must intersect the upstream CRT modules, the TPC, and the downstream CRT modules, requiring the muon to travel almost completely along the detector length axis ($Z$), again highlighted in Figure~\ref{fig:crtcoverage5758}. Therefore, the measurements presented restrict themselves to characterizing just one side of the detector, in this case, BL, where statistics are more plentiful. The CRT was only fully integrated into the data acquisition system in November 2018, restricting the number of days with data available in 2018.

\subsection{Event Selection of Cosmic-Ray Cathode-Crossing Tracks in the TPC}\label{sec:cathMethod}

Cathode-crossing cosmic-ray muons in a liquid argon TPC have been used for calorimetry calibration previously by detectors like ProtoDUNE-SP and MicroBooNE~\cite{abi2020first,Adams_2020}. These tracks cross the defined position of the cathode, providing a simple way to determine the timestamp ($\mathrm{t_0}$) of the event independent of the detector readout window~\cite{abi2020first}. With the timestamp, the reconstruction gives the tracks well-defined positions across the drift distance, enabling their usage as calibration samples in a TPC. The event selection and method are described fully in~\cite{abi2020first}, and only general details and changes to that calibration scheme will be discussed here.

\begin{figure}
    \centering
    \includegraphics[width=0.45\textwidth]{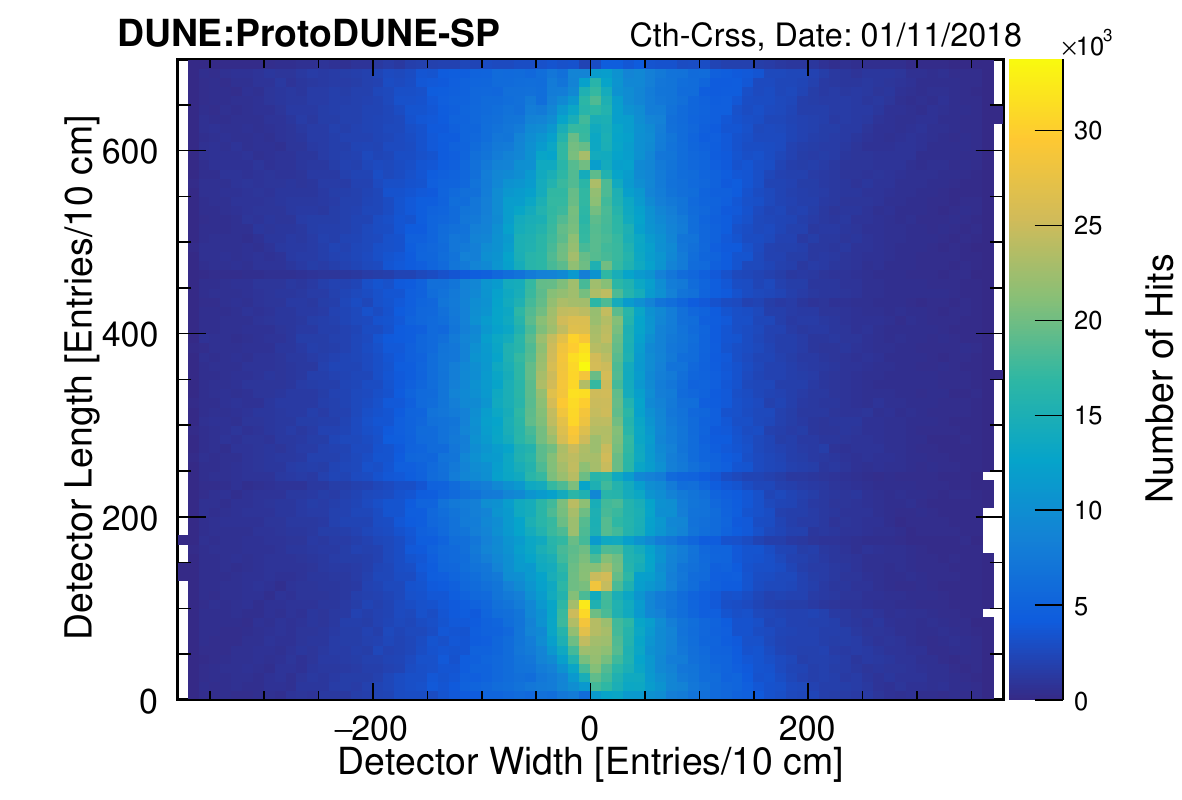}
        \includegraphics[width=0.45\textwidth]{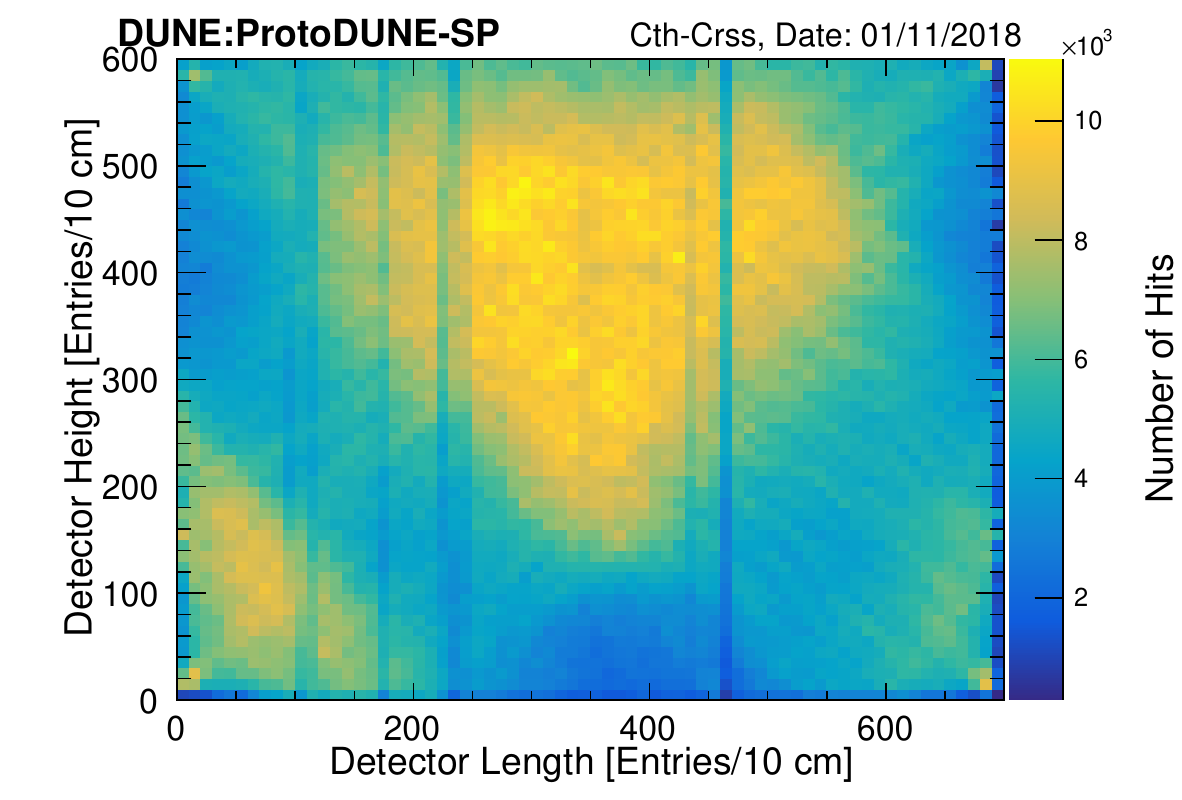}
    \caption{Statistics of hits on the collection plane collected by the cathode-crossing event selection for data taken on November 1st, 2018, from the top (left) and side (right) views. The position of hits are calibrated using the space charge effect calibration map.}
    \label{fig:cathcoverage5758}
\end{figure}

Selection steps intend to minimize distorted signals. Specifically, steps are made to ensure the trajectory of the muon is not parallel to the wire or parallel to the drift direction. Otherwise, the ionization electrons from large fractions of the whole trajectory would arrive at a single wire, making it difficult for the signal processing to form individual hints. Additionally, broken reconstructed tracks and stopping muons are eliminated. Figure~\ref{fig:cathcoverage5758} shows the statistics of signals as a function of position.


\subsection{Extracting the Drift Electron Lifetime using the TPC}

The drift electron lifetime is measured using $dQ/dx$ for drift times ranging from 0.1-2 ms binned in 0.1 ms bins. The timing window was chosen to avoid areas close to the anode or cathode at drift times of 0.0-0.1 ms and 2.0-2.3 ms, respectively. Measurements in these bins were found to be unstable, likely due to the electric field that changes near the wires and cathode.

Uncalibrated $dQ/dx$ from hits are collected in these 0.1 ms timing bins, specifically from hits on the collection plane~\cite{protoDUNEDesign}. Samples of $dQ/dx$ from the collection plane are corrected using a calibration scheme described below, and are in units of ke$^-$ per centimeter. 

An additional selection step, developed from LongBo~\cite{longbo}, is made if three successive hits associated measure charge deposited greater than 80 ke$^-$/cm, approximately the energy deposited of 2 MIPs. These trains of signals could contain delta-rays in addition to the MIP deposit from the muon signals that could bias the distributions of $dQ/dx$ measured by giving a local maximum in the tail of the distribution. Therefore, those three wire signals are discarded due to the possible delta-ray.

Corrections are applied to calibrate all effects unrelated to the drift electron lifetime. First, the space charge effect (SCE) distortions may smear the positional reconstruction for the width ($X$), height ($Y$), and length ($Z$) and are corrected. A different SCE positional correction is used for each selection to leverage the capabilities of the two event selections. For the cathode-crossing tracks, the SCE correction map is applied to the signals, which shifts the nominal positions ($X_{0}$, $Y_{0}$, $Z_{0}$) by the SCE corrections ($\Delta X_{\rm{sce}}$, $\Delta Y_{\rm{sce}}$, $\Delta Z_{\rm{sce}}$) to the corrected positions ($X_{\rm{c}}$, $Y_{\rm{c}}$, $Z_{\rm{c}}$). For CRT-TPC matched tracks, the CRT tracking information ($\Delta X_{\rm{crt}}$, $\Delta Y_{\rm{crt}}$, $\Delta Z_{\rm{crt}}$) is used as the scintillator strips are outside the detector and not susceptible to the TPC SCE effects, allowing them to calibrate positions on a track-by-track basis~\cite{abi2020first}. 

Then, the $dQ/dx_{\text{uncorr}}$ is equalized as a function of $Y$ and $Z$ using the median $dQ/dx$ across 5 cm by 5 cm voxels in the detector. The correction is applied by multiplying the $dQ/dx$ by the correction ($C_{YZ}$). The correction is the ratio of the median charge measured within that voxel in the $YZ$-plane and the median $dQ/dx$ measured for that drift volume. This step calibrates for any residual distortions in $dQ/dx$ from channel-to-channel variations and space charge effect distortions. This calibration methodology, inspired by similar work from the MicroBooNE collaboration, is discussed further in~\cite{abi2020first,Adams_2020}.

A final correction is applied to address the recombination of drift electrons in the electric field. The recombination correction ($C_{R}$) is applied as a function of the electric field at the calibrated position of a given energy deposition~\cite{abi2020first}. It assumes recombination can be parameterized using the modified Box model, with recombination constants ($\alpha$,$\beta$) as measured by ArgoNeuT (0.93, 0.212 cm/MeV)~\cite{argoneut}. The charge calibration is complete by multiplying the intermediate measurement of dQ/dx by $C_{R}$. In total, the calibrated $dQ/dx$ is obtained using the following expression:

\begin{equation}
\frac{dQ}{dx}=C_R(X_{c},Y_{c},Z_{c})C_{YZ}(Y_{c}Z_{c})\frac{dQ}{dx}_{\text{uncorr}}(X_{c},Y_{c},Z_{c}).
\end{equation}
\noindent The $dQ/dx$ is a result of removing non-uniformities in the response of electronics (YZ calibration) and applying the corrections for SCE effect, as well as recombination. Unlike in Ref.~\cite{abi2020first}, a correction in the X coordinate is not applied since that is the drift direction where the effects of the drift electron lifetime would be observed. 

\begin{figure}
     \centering
     \includegraphics[width=0.7\textwidth]{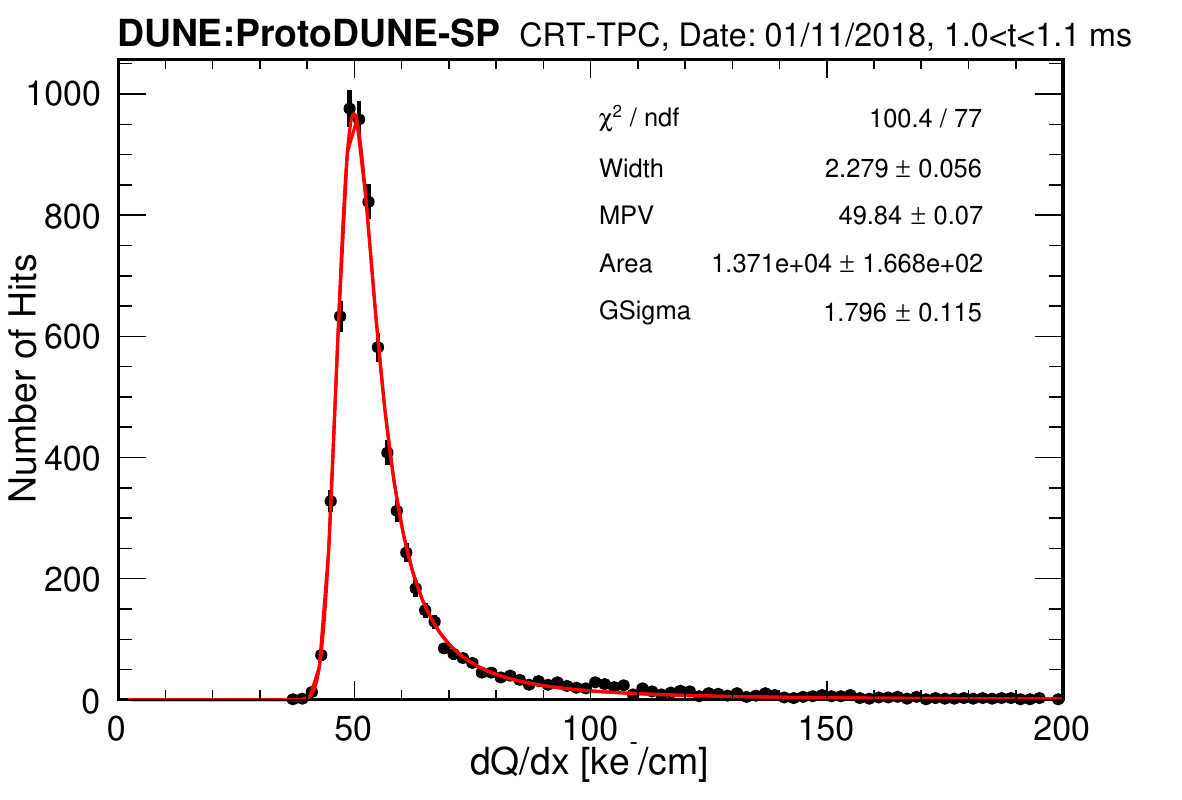}
     \caption{Example fit of a 0.1 ms timing slice from 1.0 ms to 1.1 ms using the CRT-TPC matched track selection from data taken on November 1st, 2018.}
     \label{fig:exampleLanGaus}
 \end{figure}

In each drift time bin, the most probable value (MPV) of $dQ/dx$ is measured using a fit to a Landau function convoluted with a Gaussian function (Landau-Gauss)~\cite{BRUN199781,rootRepo,langauFit} for all bins across the drift distance of the detector. A Landau distribution models energy deposits in a medium, and a Gaussian function models the detector noise, hence the use of a Landau-Gauss function. This dQ/dx has the SCE, YZ corrections, and recombination correction applied as described above. An example of a fit for one of the bins is shown in Figure~\ref{fig:exampleLanGaus}. More than 300 entries must be binned in a 0.1 ms bin, as there need to be enough statistics for a fit of the MPV.

\begin{figure}
    \centering
    \includegraphics[width=0.7\textwidth]{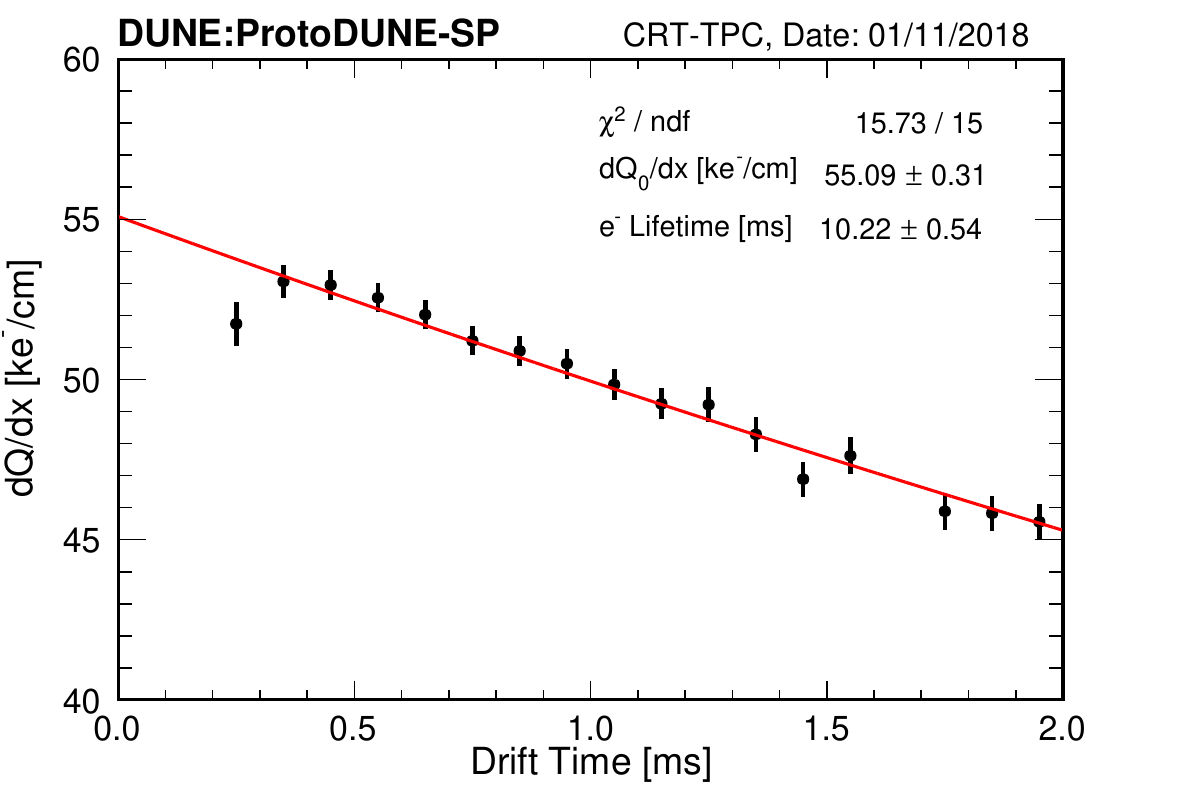}
    \includegraphics[width=0.7\textwidth]{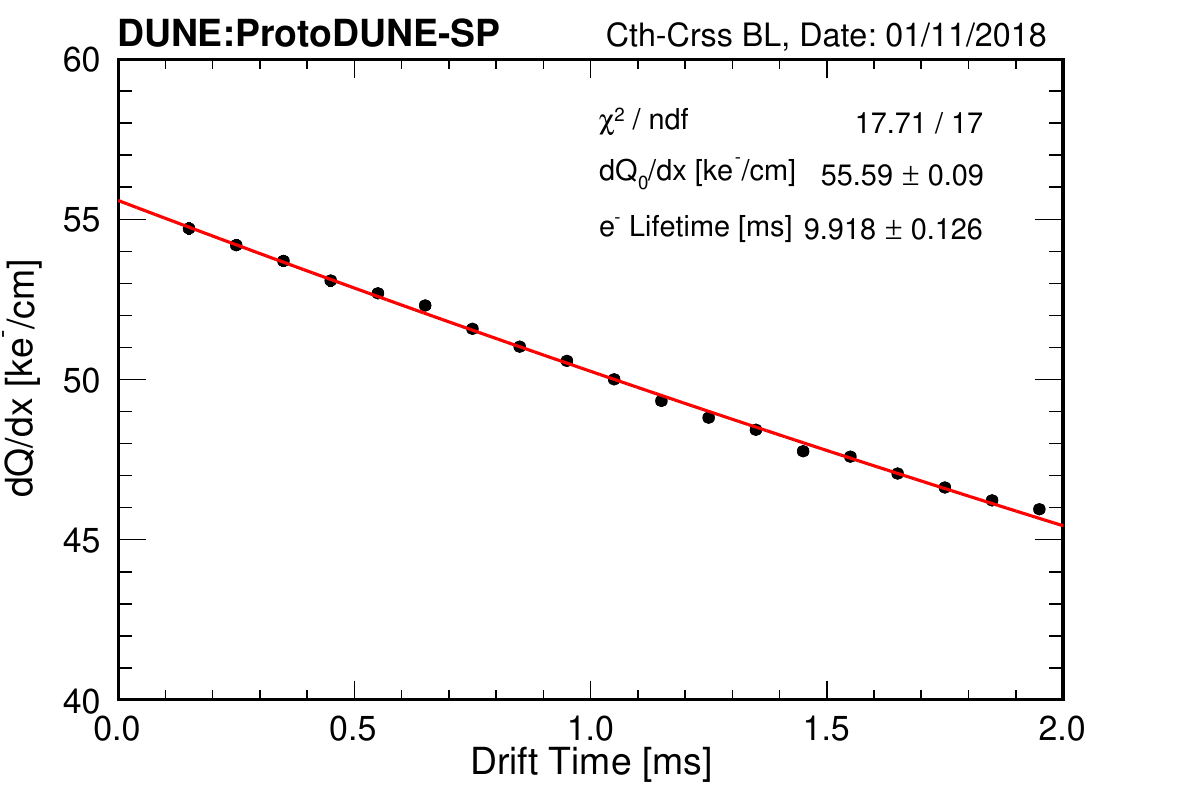}
    \caption{Drift electron lifetime measurements using the CRT-TPC selection (top) and the cathode-crossing track selection (bottom) for the same operating period on November 1st, 2018. Statistical errors, taken from the fits to the Landau-Gauss function, are used but may be too small to be visible.}
    \label{fig:exampleLifetime}
\end{figure}

The drift electron lifetime is extracted from a fit of the MPVs as a function of drift time to Equation~\ref{eqn:lifetime}, which extracts the parameters $\rm{Q_0}$ and $\tau$. The parameter $\tau$ was constrained between 0 ms and 5000 ms. Measurements are made for both the BL and BR drift volumes for the cathode-crossing sample. However, comparisons between the two TPC methods are only possible for the BL APAs given the statistics of the CRT-TPC track sample. Figure~\ref{fig:exampleLifetime} shows a drift electron lifetime measurement from the CRT-TPC matched track selection and cathode-crossing track selection for the same data run on November 1st, 2018. 

It was found that, using only the statistical uncertainty of the MPV, the statistical uncertainty for the drift electron lifetime is underestimated at low drift electron lifetimes, with the chi-square per degrees of freedom of the fit being larger than 1. Including additional systematic uncertainties to the MPV to address this was found to be inadequate as the chi-square per degrees of freedom would be less than 1 at high lifetimes, leading to an overestimation of the statistical fit uncertainty. Therefore, the statistical uncertainties on each MPV value in the fit are changed by increments of $\pm$0.001 ke$^-$/cm until the chi-square over the number of degrees of freedom is approximately 1. This procedure allowed to better estimate uncertainty of the fit parameters. The method is based on similar scaling done in Particle Data Group fits~\cite{pdg2022}. The median average scaling of MPV statistical errors per lifetime measurement is 1.10.

Measurements were made at different positions of the detector in order to test the hypothesis that liquid argon is cleaner on the BR side near the inlet, and to validate the findings of the purity monitor that the purity increases as a function of height. However, not all parts of the detector volume can be used. The front, back, top, and bottom faces of the detector can have significant space charge effect distortions. An additional buffer space is included to demarcate spaces between regions. Table~\ref{tab:volumeY} and Table~\ref{tab:volumeZ} show the boundaries of the various regions. The bottom surface of the active region of the detector represents 0 cm in height, and the front face of the active region of the detector is 0 cm across the detector length. 

\begin{table}[h]
    \centering
    \caption{Legend of the various regions of measurements taken relative to the detector height. Gaps at the top and bottom (25 cm) are to avoid SCE regions, and the gaps in the middle (50 cm) prevent measurements in a specific region from oversampling hits just on the border between regions.}
    \begin{tabular}{|c|c|c|} \hline 
   Region    & Minimum $Y$ (cm) & Maximum $Y$ (cm) \\ \hline 
     Bottom     &  25 & 175  \\ \hline 
    Middle & 225   & 375 \\ \hline 
    Top &  425   &  575  \\ \hline 
    \end{tabular}
    \label{tab:volumeY}
\end{table}

\begin{table}[h]
    \centering
    \caption{Legend of the various regions for measurements taken relative to the detector length. Labels are made relative to the position of the APAs relative to the beam, with the first set being closest to the beam and the third being the furthest away. Gaps in the regions in the upstream and downstream faces (50 cm) are to avoid places where the SCE distortions are maximal and in the middle (20 cm) to avoid distortions from signals between APAs. }
    \begin{tabular}{|c|c|c|} \hline 
    Region     & Minimum $Z$ (cm) & Maximum $Z$  (cm) \\ \hline 
    First Set of APAs     & 50 &  220 \\ \hline 
    Second Set of APAs & 240    &  450\\ \hline 
    Third Set of APAs &    470 & 645  \\ \hline 

    \end{tabular}
    \label{tab:volumeZ}
\end{table}

It is instructive to consider argon purity in terms of the fraction of the charge remaining after it drifts the full distance from cathode ($Q_c$) to anode ($Q_a$). The largest charge attenuation corresponding to a maximum drift time (2.25 ms) for a central-value, $\tau_{CV}$, of the measured drift electron lifetime is given by:

\begin{equation}
\rm{\frac{Q_a}{Q_c}=\exp({-2.25 \, ms/\tau_{CV}})}.
\label{eqn:qaqccv}
\end{equation}

The formulation of Equation~\ref{eqn:qaqccv} gives a direct figure of merit for the minimal fraction of the charge remaining for a given value of the drift electron lifetime. That is to say it is the fraction of the charge that survives if the ionization charge cloud were to traverse the entire drift volume from the cathode to the anode. To quantify the impact on the calorimetric measurements in the TPC, one can parametrize the uncertainty ($\sigma_Q$) on the measured charge ($Q$) due to presence of impurities as:

\begin{equation}
\rm{\frac{\sigma_Q}{Q}(t)=\frac{t \sigma_{\tau}}{\tau^2}}.
\label{eqn:qfracuncert}
\end{equation}
\noindent As seen in Equation~\ref{eqn:qfracuncert}, the uncertainty on the charge measurement decreases quickly as the drift lifetime increases, as it is proportional to $1/\tau^2$. 

\subsection{Systematic Uncertainties for the Drift Electron Lifetime Using the TPC}

The three systematic uncertainties address the electric field corrections, recombination parameters used, and diffusion of drifting ionization electrons. These uncertainties address the modeling of $dQ/dx$ and the modeling of the detector's electric field. 

The first uncertainty involves the electric field corrections. These corrections come from the data-driven SCE map that measures the positional distortions on the surface of the detector and uses that to scale and measure an electric field map correction~\cite{abi2020first}. The space charge effect uncertainty is evaluated by applying the alternative correction map to the recombination correction, which depends on the local electric field. The alternative map determines the electric field distortions by measuring positional offsets at the crossing points of muon tracks. The difference between measurements using the nominal and alternative correction map is used as the uncertainty. The re-measured drift electron lifetime is labeled as $\tau_{\text{alt  sce}}$. 

The second set covers the uncertainty from the recombination factors, specifically the model parameters. It takes the uncertainty on the $\alpha$ and the $\beta$ parameters from the ArgoNeuT analysis on recombination~\cite{argoneut}. It repeats the measurements with these altered parameters for the recombination correction. The $\alpha$ and $\beta$ parameters are shifted in the same direction, maximizing the shifts' impacts. The uncertainty is acquired by calculating the difference between these measurements and the central value for each data point.

The third uncertainty is tied to the fact that diffusion may alter the drift electron lifetime measurement. Electron diffusion affects the reconstructed dQ/dx~\cite{Putnam:2022lli}. We have not applied corrections for the diffusion and estimated the potential bias by finding difference in estimates of the lifetime in simulations without diffusion and with diffusion using nominal diffusion constants of 12.8 cm$^2$/s for transverse diffusion and 5.3 cm$^2$/s for longitudinal diffusion. The whole difference between the measurements is treated as the associated uncertainty. Diffusion constants in liquid argon themselves have large uncertainties~\cite{uBooNeDrift}; therefore, the treatment is considered necessary given previous experience with ProtoDUNE-SP drift electron analyses~\cite{abi2020first}. The differences between the measurements of $\frac{Q_a}{Q_c}$ from simulations with and without diffusion are 2.8\% using cathode-crossing tracks and 3.6\% for CRT-TPC matched tracks. The central-value $\frac{Q_a}{Q_c}$ for each measurement is shifted up and down by these values for each event selection, respectively.

The ranges of measurements, including the systematic uncertainties, with $\rm{\tau_{CV}}$ and $\rm{\sigma_{stat.}}$ representing the central-value measurement and statistical fitting uncertainty, are defined below for the upper bound ($\sigma_{\tau \, high}$) and lower bound ($\sigma_{\tau \, low}$) in Equation~\ref{eqn:high} and Equation~\ref{eqn:low}, respectively. The systematic errors are conservatively added linearly. In these equations, $\rm{\tau_{alt \, sce}}$ represents changing the space charge effect map, and $\rm{\tau_{\pm1\sigma \, recomb}}$ represents altering the recombination parameters. The value $\rm{\sigma_{+1\sigma \, diff}}$ represents the diffusion uncertainty translated into units of drift electron lifetime for a given $\rm{\tau_{CV}}$ measurement. The full errors are then propagated for both the high band and low band of the full uncertainties for $\rm{\frac{Q_a}{Q_c}}$ in Equation~\ref{eqn:qaqcErr}. The formula comes from taking the derivative of Equation~\ref{eqn:qaqccv}.

\begin{equation}
\rm{\sigma_{\tau \, high}=\sqrt{\sigma_{stat}^2+(|\tau_{CV}-\tau_{-1\sigma \, recomb}|+|\tau_{CV}-\tau_{alt \, sce}|+\sigma_{+1\sigma \, diff})^2}}
\label{eqn:high}
\end{equation}

\begin{equation}
\rm{\sigma_{\tau \, low}=\sqrt{\sigma_{stat}^2+(|\tau_{CV}-\tau_{+1\sigma \, recomb}|+|\tau_{CV}-\tau_{alt \, sce}|+\sigma_{-1\sigma \, diff})^2}}
\label{eqn:low}
\end{equation}

\begin{equation}
\rm{\frac{Q_a}{Q_c}_{high/low}=\frac{Q_a}{Q_c}_{CV}\pm\frac{Q_a}{Q_c}_{CV}\frac{2.25 \, ms \times \sigma_{\tau \, high/low}}{\tau^2}}
\label{eqn:qaqcErr}
\end{equation}

\subsection{Analysis with the CRT-TPC Selection and the Cathode-Crossing Selection}

\begin{figure}[h]
    \centering
    \includegraphics[width=0.8\textwidth]{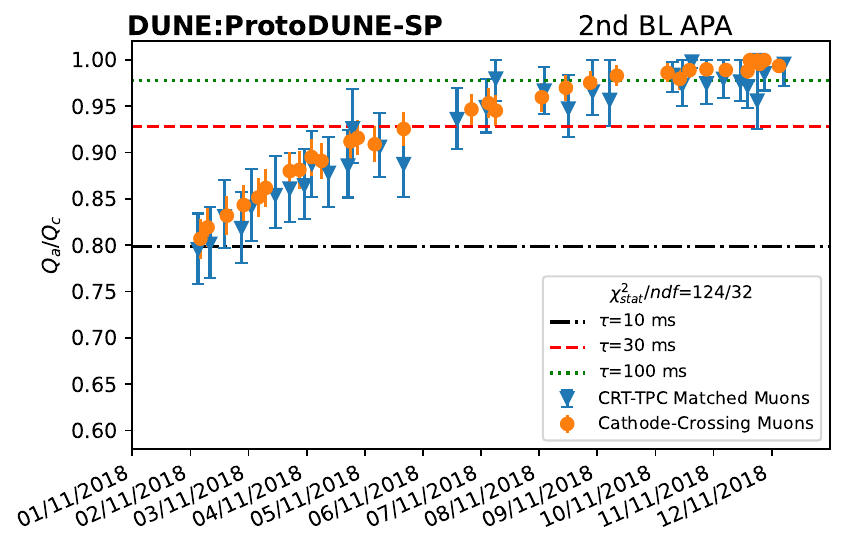}
    \caption{The ratio of minimum charge remaining from cathode to anode using the drift electron lifetime measurements with the CRT-TPC track selection and the cathode-crossing track selection. Both statistical and systematic errors are shown. However, the chi-square statistic between the two datasets is calculated only using the errors from the statistical fit of the drift electron lifetime.}
    \label{fig:crtCompare}
\end{figure}

The drift electron lifetime measurements were taken at various positions and heights. The first assumption tested was if the drift electron lifetime measurements at the middle height of the detector for signals on the second set of APAs agreed between the cathode-crossing selection and the CRT-TPC selection since this is the only volume in which statistics are plentiful enough for the latter. Figure~\ref{fig:crtCompare} shows the comparisons during runs in which the CRT was active on November 1st, 2018, with both event selections producing measurements have the same trend within the statistical and systematic uncertainties. The comparison verifies that the cathode-crossing event selection is suitable for extracting the liquid argon purity at other detector locations. These time series are plotted with a reference line to the drift electron lifetime of 10 ms, the technical specification of the DUNE Horizontal Drift module. However, the data clearly far exceeds this reference value and this reference line will be omitted for proceeding comparisons. A reduced chi-square statistic is calculated based on the nearest data points and uncertainties between the data sets. The data points must be within six hours of each other to be considered for the chi-square statistics to ensure they probe the purity at a similar time. For comparisons between the two TPC methods, only the statistical uncertainties are used as the methods for obtaining the systematic uncertainties are shared between measurements using both samples of TPC tracks.

\begin{figure}
    \centering
        \includegraphics[width=0.65\textwidth]{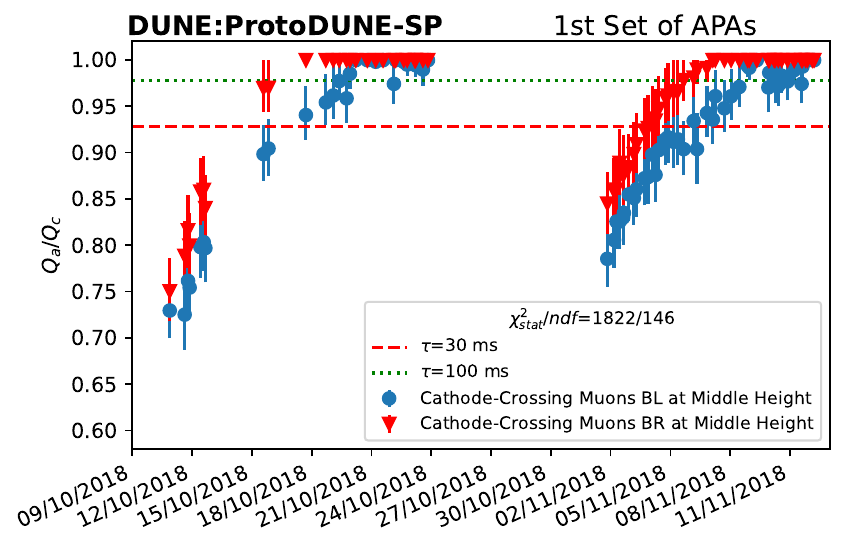}
    \includegraphics[width=0.65\textwidth]{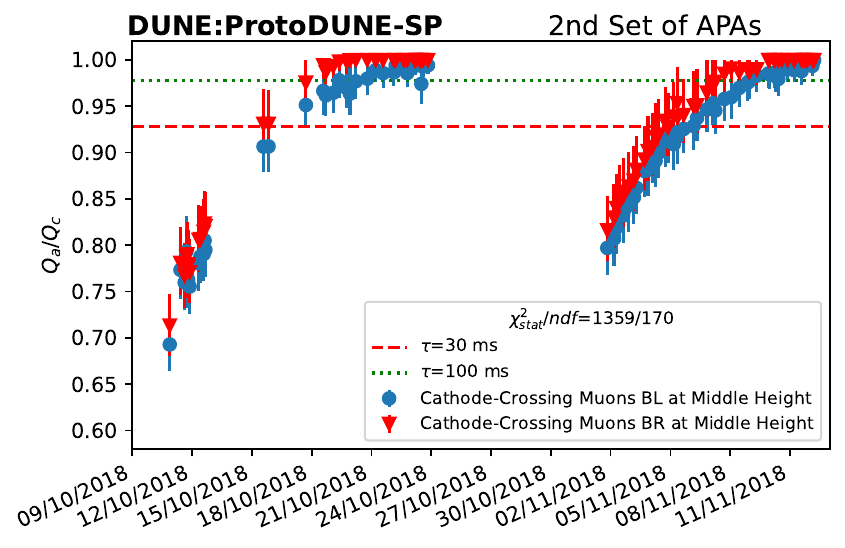}
        \includegraphics[width=0.65\textwidth]{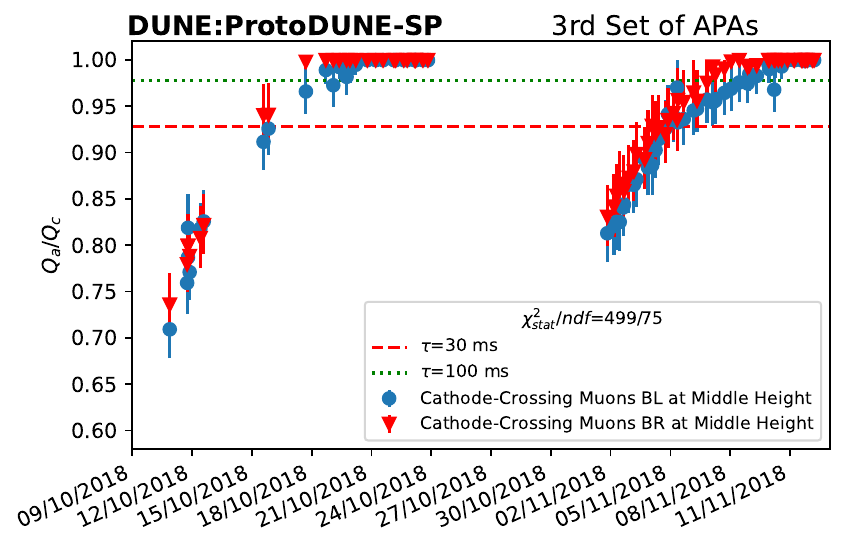}
    \caption{Comparisons of the minimum fraction of charge remaining from drift electron lifetime measurements between the BL and BR drift volumes for each set of APAs with the cathode-crossing muon selection. Both statistical and systematic errors are shown. The chi-square statistic between the two datasets is calculated with only the statistical fit errors for measuring the drift lifetime.}
    \label{fig:BLvsBR}
\end{figure}

As mentioned in Section~\ref{sec:pdsp}, the BL side contains the liquid argon outlet, and the BR side has the liquid argon inlet. Therefore, a proper validation test is that the BR side should have a higher charge remaining than the BL side. In Figure~\ref{fig:BLvsBR}, the BR side's central-value measurements are consistently higher.

These tests showed that the drift electron lifetime measurements were above 30 ms for an extended period in 2018, well above the 10 ms technical specification of the DUNE Far Detector Horizontal Drift module. For weeks, the lifetime achieved is infinite, notably from October 16th to October 22nd, after which a drop in purity occurred due to a stop in the recirculation pump~\cite{protoDUNEDesign}. The comparison also validates the usage of the cathode-crossing event selection. This may not appear obvious, given the high chi-square disagreement. However, the two methods follow each other to high lifetimes in November. The drift electron lifetime measurements need to be high for low uncertainties on energy calorimetry, so calibrations at high lifetimes do not degrade the performance of the calorimetry, regardless of discrepancies between methods. Lastly, the measurements with the cathode-crossing samples in Figure~\ref{fig:BLvsBR} validate assumptions about the liquid argon purity based on the placement of cryogenic instrumentation, whereby cleaner argon comes in from the bottom BR, and less pure argon leaves the detector BL. 

More importantly, the purity is high and reaches drift electron lifetimes above 100 ms. These high lifetimes have systematic uncertainties for $Q_a/Q_c$ at the 1\%-level, equating to an uncertainty on the calorimetry due to impurities at the sub-percent level. Both plots highlight the capability of the DUNE design and the negligible uncertainties observed by ProtoDUNE-SP for calibrating the impacts of impurities on the calorimetry.

\section{Measuring the Drift Electron Lifetime using Purity Monitors}\label{sec:PrMon}

A purity monitor is a miniature TPC that measures the lifetime with photoelectrons generated from its UV-illuminated gold photocathode based on designs from ICARUS~\cite{icarus}. Three purity monitors were installed in the ProtoDUNE-SP detector, which were used to independently infer the free electron lifetime in the detector. The UV light is produced by an external xenon light source. The light then travels via quartz fibers to the cathode, leading to photoelectrons in the purity monitor drift chamber. The source can produce light at a maximum repetition rate of 100 Hz with a wavelength of approximately 250 nm. Assuming a uniform electric field inside the purity monitor volume, the fraction of photoelectrons arriving at the anode of the purity monitor is a measure of the electron lifetime: 
\begin{equation}
Q_\mathrm{a}/Q_\mathrm{c}=e^{-t/\tau},
\label{eqn:prm_qratio}
\end{equation}
where $Q_\mathrm{c}$ and $Q_\mathrm{a}$ are the charge of electrons generated at the cathode and collected at the anode, respectively. There are multiple flashes from the light source for each measurement. Therefore, the results have small statistical errors. In addition, the space charge effect caused by cosmic rays is negligible in a purity monitor, given its small active volume compared to the large TPC volume~\cite{Palestini:2021mlc}. The purity monitors have been proven to provide quick, reliable, real-time information during detector commissioning~\cite{protoDUNEDesign}. Furthermore, they have unique importance for the charge calibration of DUNE's Far Detector, where cosmic ray-based calibration becomes challenging due to low statistics.

\subsection{Design of the Purity Monitor}
\begin{figure}
    \centering
    \includegraphics[width=0.7\textwidth]{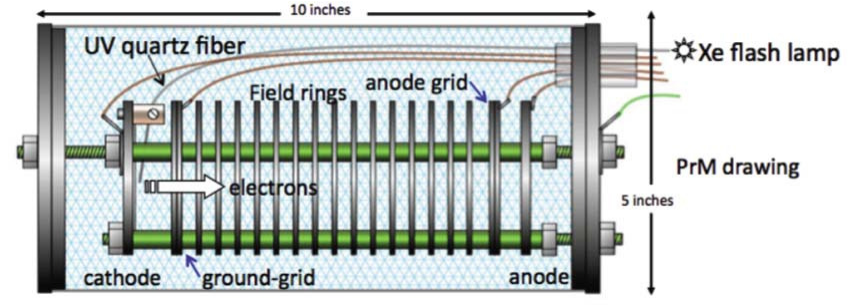}
    \caption{A conceptual diagram of the purity monitor used in ProtoDUNE-SP. The image is from~\cite{Adamowski:2014daa}.}
    \label{fig:purMon}
\end{figure}

The three purity monitors were installed at heights of 1.8 m, 3.7 m, and 5.6 m from the bottom of the ProtoDUNE-SP cryostat. Each consists of four parallel, circular electrodes: a cathode disk holding a photocathode, an anode disk, and two grid rings, one in front of the anode and the other in front of the cathode. A schematic diagram of a purity monitor is shown in Figure~\ref{fig:purMon}. The distances between cathode and cathode-grid, between the grids, and between anode-grid and anode are referred to as $d_1$, $d_2$, and $d_3$, respectively, and are shown in Table~\ref{tab:PrM_lengths}. The top and bottom purity monitors are constructed to be the same length. However, the middle purity monitor has a slightly shorter $d_2$.  
The cathode-grid is connected to the ground potential. The cathode and anode of each purity monitor can be biased to adjust the strength of the electric field in the drift volume. The transparency of the grid rings to free electrons is determined by the geometry of the grids and the ratio of electric field strength at the two sides of the grid rings~\cite{bunemannDESIGNGRIDIONIZATION2011}. In between two grid rings, a chain of coaxial rings is interconnected by resistors to ensure the uniformity of the electric field. The measured resistance between the two grid rings $R_g$ is shown in Table~\ref{tab:PrM_lengths}. In addition, a 110 M$\Omega$ resistor exists between the anode-grid and the anode. These resistors divide the potential between the anode and the cathode-grid based on the resistance. For the geometry of the purity monitors used in this detector, the grid is electrically transparent to the drifting electrons when the absolute ratio of bias voltages applied to the anode ($V_a$) and the cathode ($V_c$) is greater than 20.

\begin{table}[h]
    \centering
    \caption{The distances between cathode and cathode-grid ($d_1$), between the grids ($d_2$), and between anode-grid and anode ($d_3$). The total resistance between the cathode-grid and the anode-grid ($R_g$) is also shown.}
    \begin{tabular}{|c|c|c|c|c|} \hline 
    Label     & $d_1$ (cm) & $d_2$ (cm) & $d_3$ (cm) & $R_{g}$ (M$\Omega$)\\ \hline 
     Bottom     &  1.9 & 15.74 & 0.88 & 747 \\ \hline 
    Middle & 1.9   &  14.75 & 0.88 & 696 \\ \hline 
    Top &  1.9   &  15.74  & 0.88 & 761 \\ \hline 
    \end{tabular}
    \label{tab:PrM_lengths}
\end{table}

From the experience of previous LArTPC experiments, low signal strength has limited the precision and measuring range of purity monitors. In ProtoDUNE-SP, the UV light was delivered to each purity monitor by eight fibers and an 8-channel feedthrough. It enhances the signal magnitude by a factor of six compared to the signal from a single fiber. Given $Q_\mathrm{a}/Q_\mathrm{c}=e^{-t/\tau}$, the relative uncertainty in lifetime $\Delta\tau/\tau$ is propagated as $\Delta\tau/\tau=(\tau/t)\frac{\Delta(Q_\mathrm{a}/Q_\mathrm{c})}{(Q_\mathrm{a}/Q_\mathrm{c})}$. The formula shows the relative uncertainty in the measured lifetime is proportional to the electron lifetime itself $\tau$ and inversely proportional to the drift time $t$. Therefore another improvement of the sensitivity was obtained by controlling the electron drift time, which depends on the cathode-anode distance and the applied voltage. During operation, the purity monitors were set at low anode-to-cathode voltage ratios ($|V_a$/$V_c|$), which allowed electrons to move with longer drift time and a larger difference between the charge collected at the cathode and anode. It eventually reduced the uncertainty of the measured electron lifetime compared to the configuration with a shorter drift time.

\subsection{Purity Monitor Data}

\begin{figure}[htbp!]
    \centering
    \includegraphics[width=0.48\textwidth]{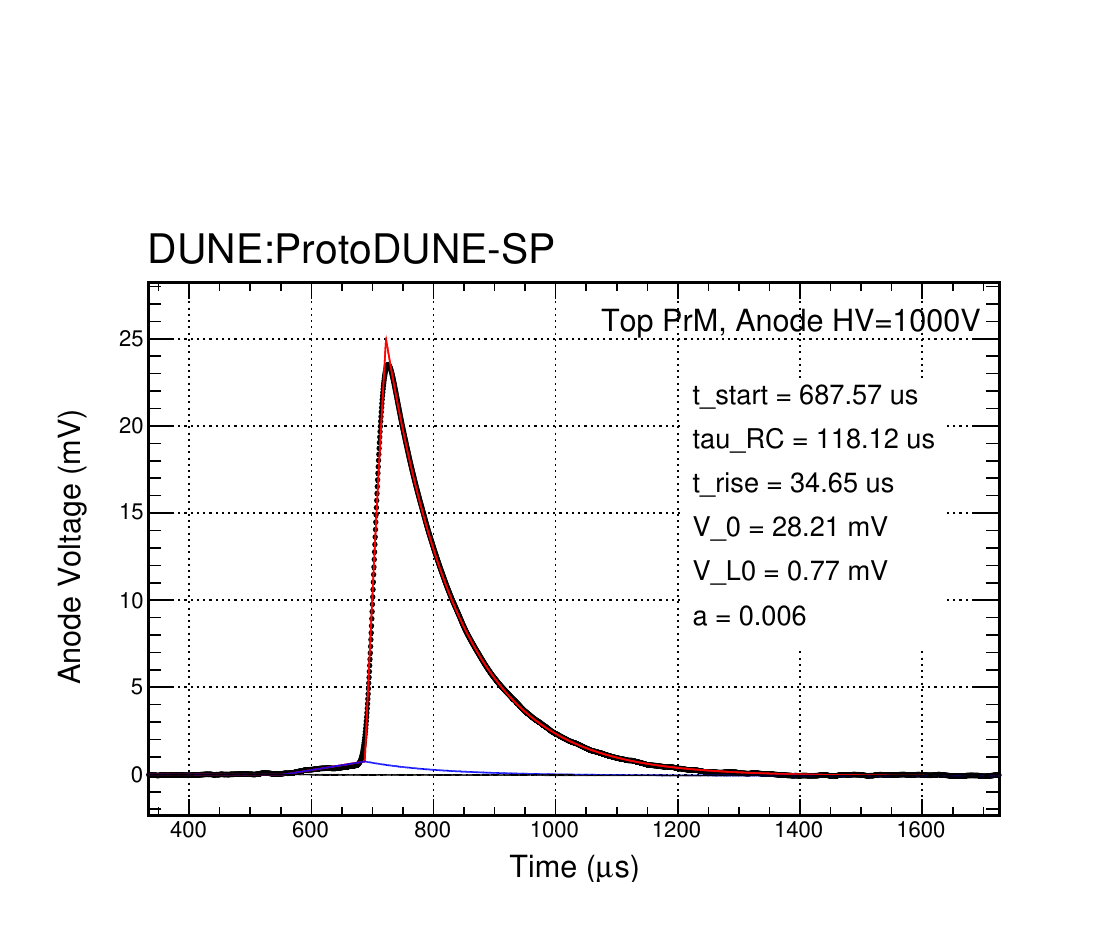} \quad
    \includegraphics[width=0.48\textwidth]{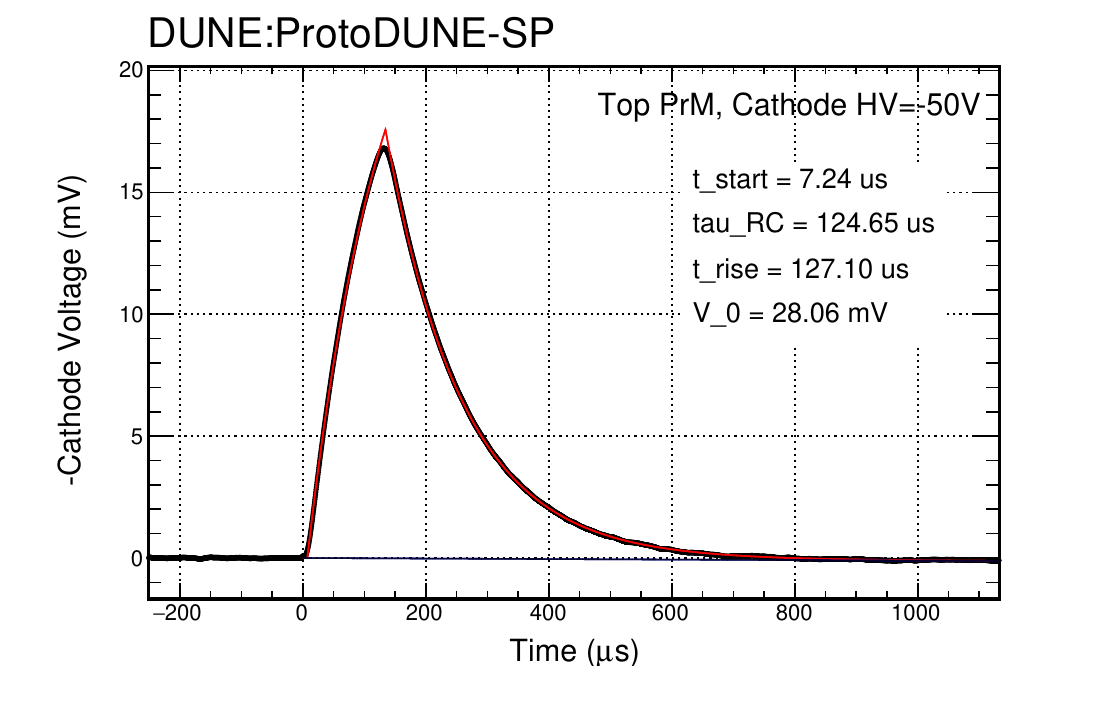}
    \caption{An example of the waveform from the anode (left) and cathode (right) of the top purity monitor. The red and blue curves represent the functional fits of the purity monitor data (black).}
    \label{fig:prm_waveform}
\end{figure}

Each electron-lifetime measurement by the purity monitors is based on the signal ratios $Q_\mathrm{a}/Q_\mathrm{c}$ from 200 UV flashes at the same location within 40 seconds. The signals induced on the cathode and anode are fed into two charge amplifiers in a purity monitor electronics module. The charge amplifiers have a 10 pF integration capacitor with a 22 M$\Omega$ resistor in parallel with the capacitor. A 2 MS/s, 12-bit digitizer read the amplified signals. The waveforms are averaged for each measurement to mitigate fluctuations. In addition, a special run was taken with zero voltages applied to characterize the baseline of the electronics and external noise. This measured baseline was subtracted from the raw waveform to correct the rising and falling edges. An example of the baseline-subtracted waveform from the anode of the top purity monitor is shown in Figure~\ref{fig:prm_waveform}. The maximum voltage is distorted due to the RC circuits of the charge amplifiers, and a correction with rise time needs to be applied to obtain the actual signal. A model is established to parameterize the waveform and extract the charge:

\begin{align*}
	\mathrm{Rising~edge} &: V(t)  =V_0\frac{1-\exp(-t/\tau_\mathrm{RC})}{t_{\mathrm{rise}}/\tau_\mathrm{RC}},  \\
	\mathrm{Observed~maximum~voltage} &: V_{\mathrm{max}}  =V(t_{\mathrm{rise}})=V_0\frac{1-\exp(-t_{\mathrm{rise}}/\tau_\mathrm{RC})}{t_{\mathrm{rise}}/\tau_\mathrm{RC}}, \\
	\mathrm{Falling~edge} &: V(t) =V_{\mathrm{max}}\exp(-\frac{t-t_{\mathrm{rise}}}{\tau_\mathrm{RC}}),
\end{align*}

\noindent where $t$ is the time span after the start $t_{\mathrm{start}}$, $t_{\mathrm{rise}}$ is the rising time, $V_0=Q/C$ is the signal from the purity monitor. $t_{\mathrm{start}}$, $t_{\mathrm{rise}}$,  and $V_0$ are the three free parameters in the fit. $\tau_\mathrm{RC}$ is the time constant of the circuit and is measured in-situ by fitting the falling edge of the waveform. There is an extra term for the anode signal, as shown in blue in Figure~\ref{fig:prm_waveform}. It happens when the anode-grid cannot completely shield the anode from the fields induced by the moving electrons in the drift volume, which is characterized by the inefficiency of the grid~\cite{bunemannDESIGNGRIDIONIZATION2011}. In the waveform model, a linear term is used to describe the rising of voltage before electrons pass through the anode-grid, and an exponential term is used for the falling of voltage after that point:

\begin{align*}
	V_L(t) &= V_{L0}+at,~t<0,	\\
	V_L(t) &= V_{L0}\exp(-\frac{t}{\tau_\mathrm{RC}}),~t>0.
\end{align*}

As shown in Figure~\ref{fig:prm_waveform}, this model (red line) describes well the purity monitor data (black line). The anode has a shorter rising time than the cathode due to a strong electric field present between the anode and the anode-grid, compared to that between the cathode-grid and the cathode.

\begin{figure}[htbp!]
    \centering
    \includegraphics[width=0.45\textwidth]{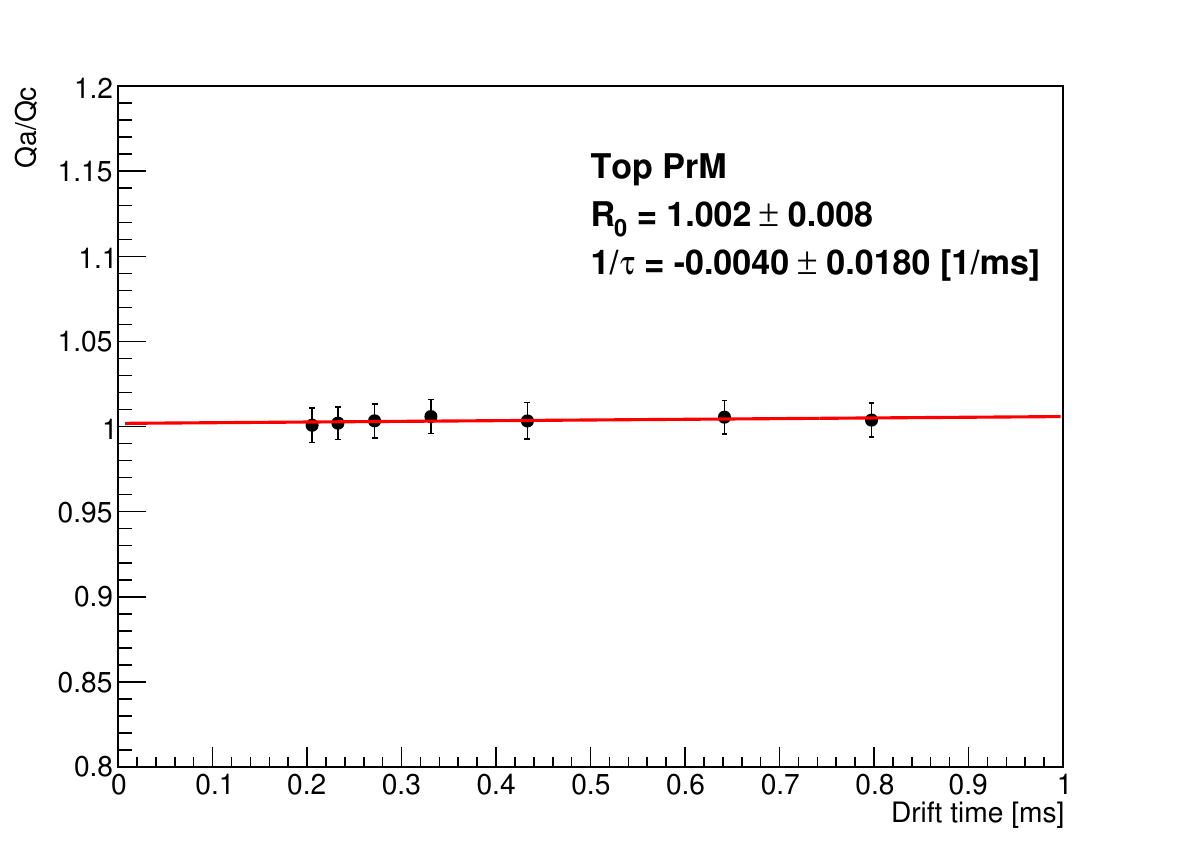}
    \includegraphics[width=0.45\textwidth]{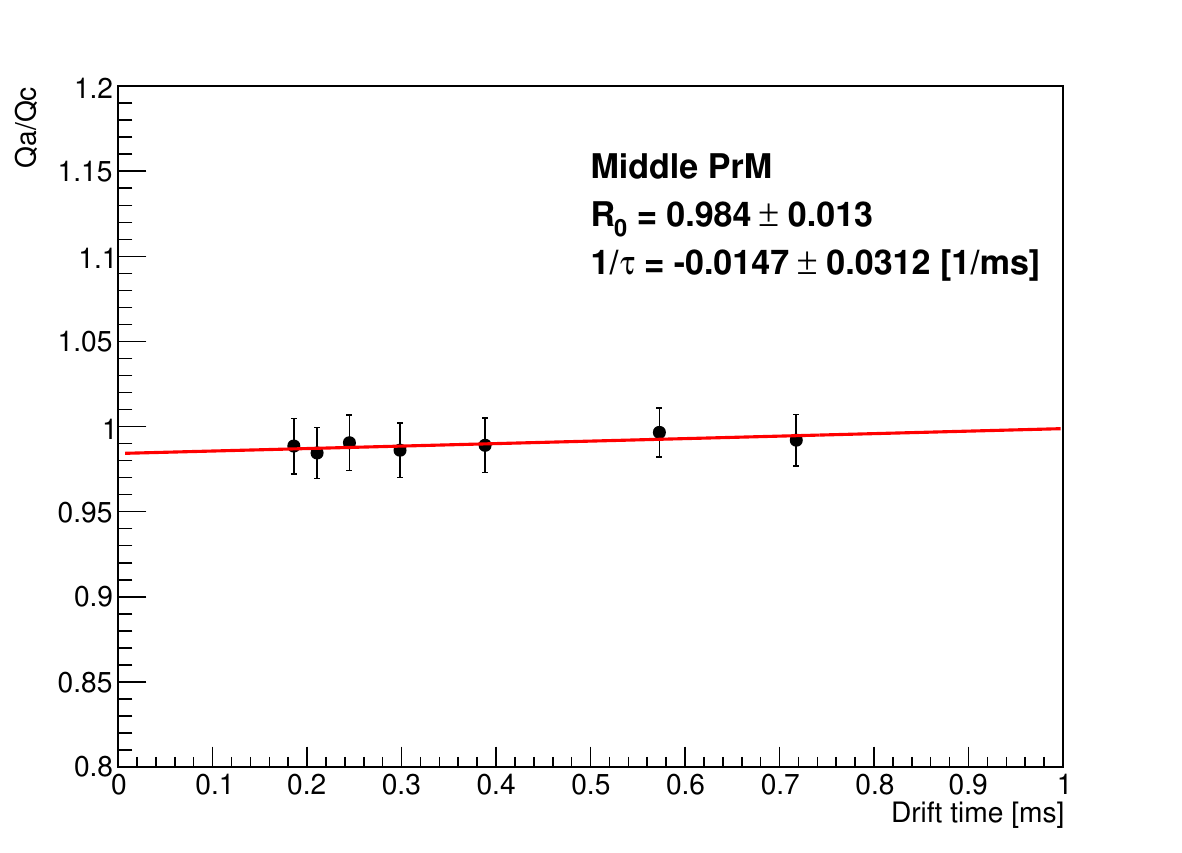}
    \includegraphics[width=0.45\textwidth]{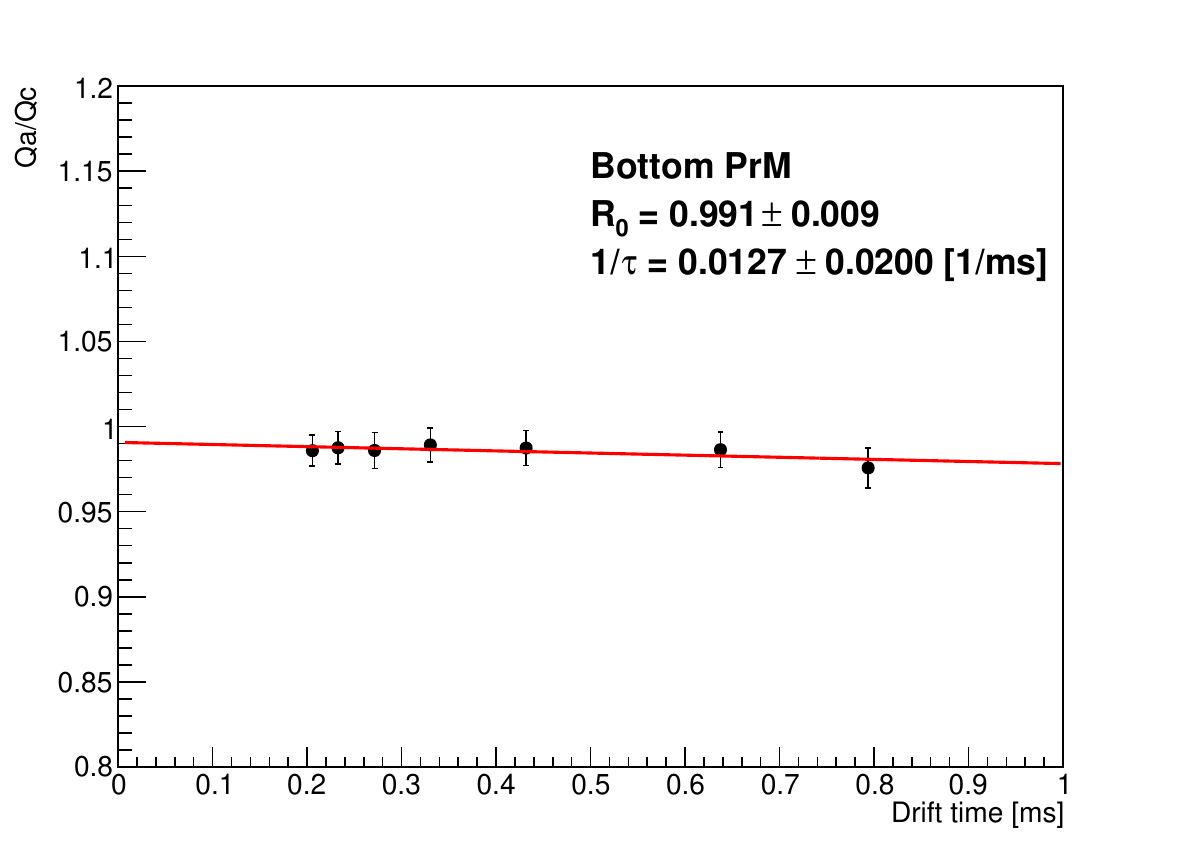}
    \caption{Measured $Q_\mathrm{a}/Q_\mathrm{c}$ at different drift times obtained by using seven calibration points for the top, middle, and bottom purity monitors. The red line is the fit with Equation~\ref{eqn:prm_qratio} to the calibration points.}
    \label{fig:prm_transparency_correction}
\end{figure}

The purity monitors were normally operated at anode-to-cathode voltage ratios of ($V_a$/$V_c$: 250 V/-50 V or 500 V/-50 V), corresponding to the field strengths in the main drift volume of about 15 V/cm and 30 V/cm, respectively. Correspondingly, the electrons drift over the volume in about 2.3 ms and 1.2 ms. The long drift time effectively lowers the charge ratio and reduces the uncertainty in the measured lifetime. However, the transparency of the cathode-grid to electrons also diminishes at such anode-to-cathode voltage ratios. A transparency correction was needed when calculating the actual charge ratio,
\begin{equation}
	\left(\frac{Q_\mathrm{a}}{Q_\mathrm{c}}\right) ^{\mathrm{corrected}}_{250/-50}=\left(\frac{Q_\mathrm{a}}{Q_\mathrm{c}}\right)_{250/-50}\times f_\mathrm{trans},
\label{eq:correctedQratio}
\end{equation}
where $f_\mathrm{trans}$ is the transparency correction factor. Dedicated calibration measurements with maximum transparency configurations were carried out at: 800 V/-40 V, 1000 V/-50 V, 1500 V/-75 V, 2000 V/-100 V, 2500 V/-125 V, 3000 V/-150 V, 3500 V/-175 V. The drift times of the calibration runs range from 0.2 to 0.8 ms. Figure~\ref{fig:prm_waveform} shows an example of the waveforms from the top purity monitor taken at the calibration point (1000 V/-50 V). The $Q_\mathrm{a}/Q_\mathrm{c}$ with full transparency at different drift times obtained from the calibration points is shown in Figure~\ref{fig:prm_transparency_correction}. Equation~\ref{eqn:prm_qratio} was used to fit the calibration points to extract the lifetime and to calculate the expected $Q_\mathrm{a}/Q_\mathrm{c}$ with the drift time at the normal operation voltages. In addition to the lifetime $1/\tau$, the $\mathrm{R}_0$ parameter was added as an overall scale parameter. $\mathrm{R}_0$ is able to float in the fit to account for the gain difference between the cathode and anode. The transparency correction factor is determined by: 
\begin{equation}
	f_\mathrm{trans} = \left(\frac{Q_\mathrm{a}}{Q_\mathrm{c}}\right) ^{\mathrm{expected}}\bigg/\left(\frac{Q_\mathrm{a}}{Q_\mathrm{c}}\right)^{\mathrm{observed}}.
\label{eq:transparencyCorrection}
\end{equation}

\subsection{Electron Lifetime Measurements from the Purity Monitors}
\begin{figure}[htbp!]
    \centering
    \includegraphics[width=\textwidth]{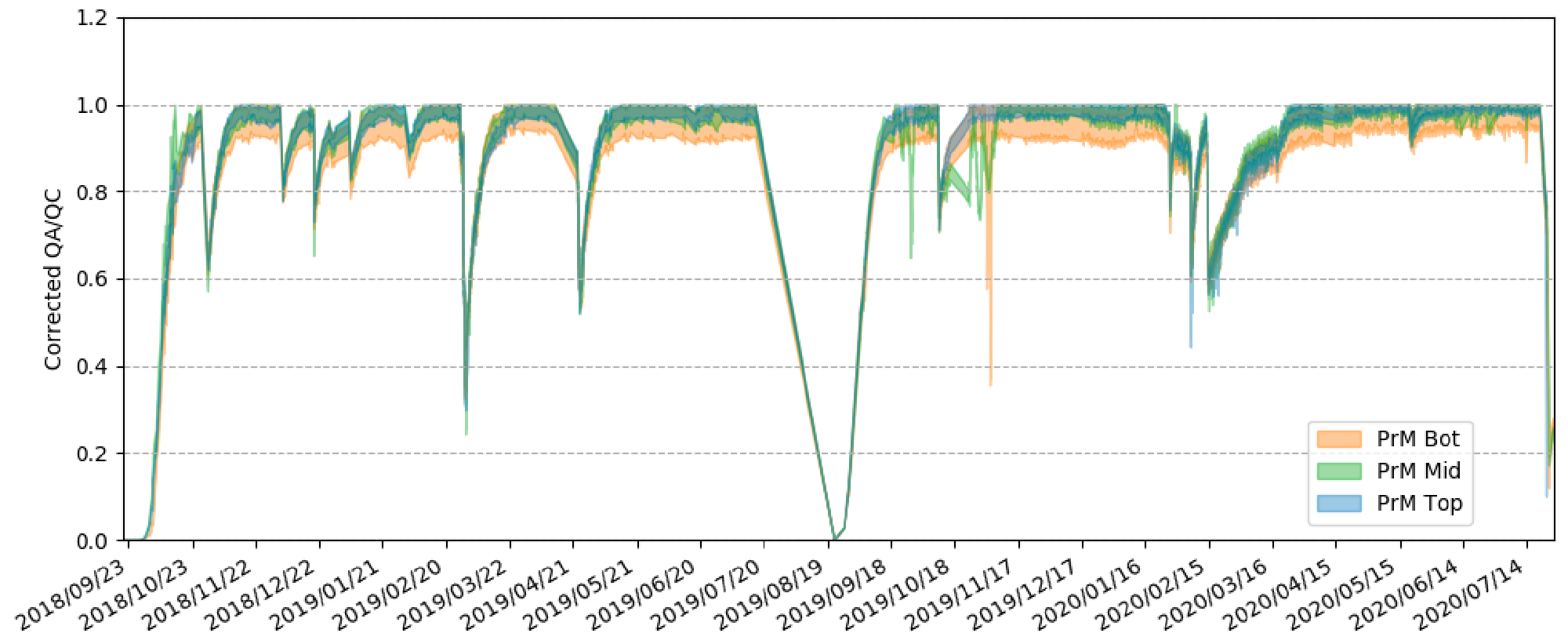}
    \caption{The anode-to-cathode signal ratios $Q_a/Q_c$ measured by three purity monitors as a function of time, September 2018 through July 2020. The purity was low before the start of circulation in October 2018. Later dips represent recirculation studies and recirculation pump stops. The shaded bands represent uncertainties of the measurements. The image is originally from~\cite{protoDUNEDesign}.}
    \label{fig:prm_correctedQratio}
\end{figure}

The corrected $Q_\mathrm{a}/Q_\mathrm{c}$ as shown in Figure~\ref{fig:prm_correctedQratio} were measured from the commissioning phase, which started in September 2018, through the entire beam test, which ended in November 2018, and continued through the non-beam phase, which ended in February 2020. The shaded bands represent the uncertainties in the measurements. The systematic uncertainties address corrections in the baseline subtraction, RC constants, the inefficiency of the grid shielding, and the transparency correction, as well as changes in the gain of the purity monitor over time. The overall uncertainties in the charge ratio measurement $\frac{\Delta(Q_\mathrm{a}/Q_\mathrm{c})}{(Q_\mathrm{a}/Q_\mathrm{c})}$ at 2.3 ms drift time are 1.9\%, 2.2\%, and 3.9\% for the top, middle, and bottom purity monitors, respectively. Uncertainties on the gain of the device and the transparency corrections equally dominate these uncertainties. 


\section{Comparisons with the TPC and Purity Monitor Measurements}\label{sec:compare}

\begin{figure}
    \centering
    \includegraphics[width=0.65\textwidth]{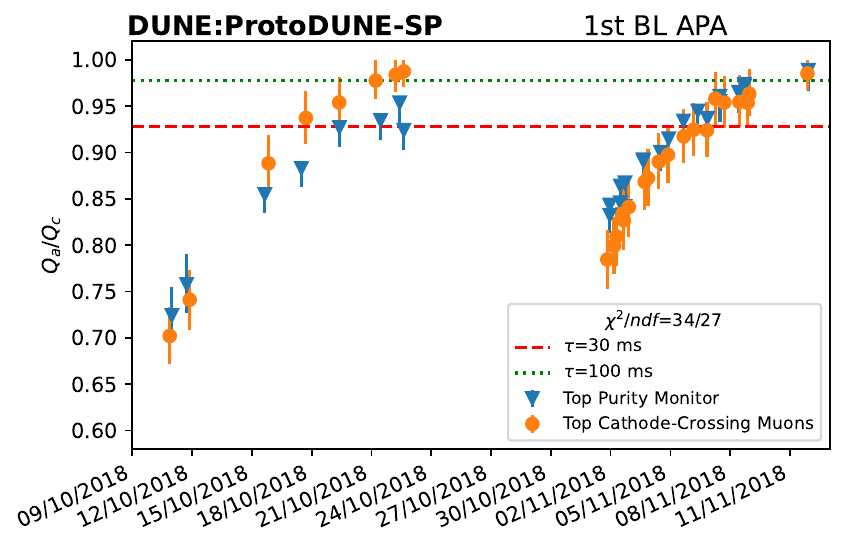}
        \includegraphics[width=0.65\textwidth]{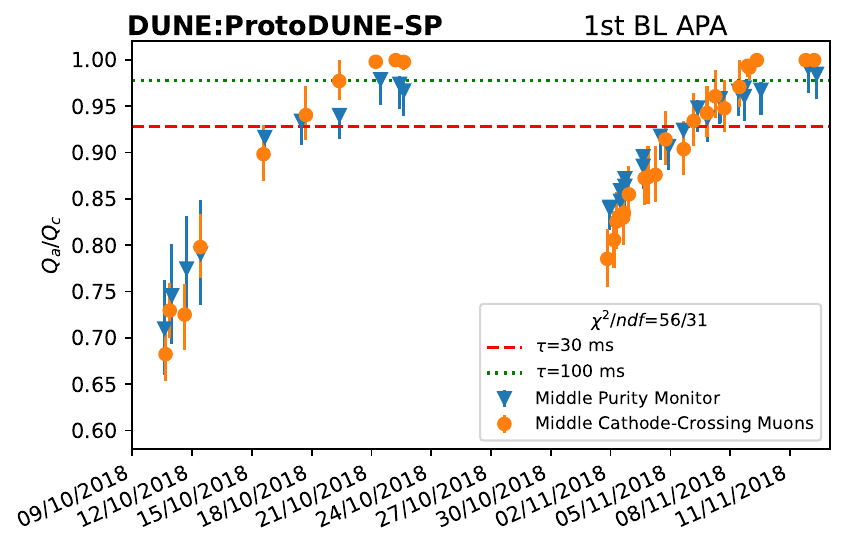}
    \includegraphics[width=0.65\textwidth]{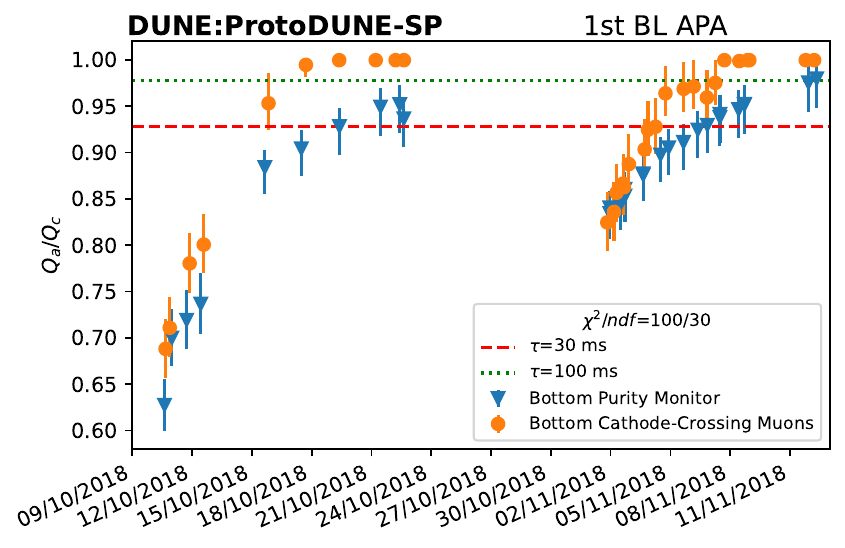}
    \caption{Measurements of the minimum fraction of charge remaining from cathode-crossing muons in the TPC and the purity monitors. Each height represents a different purity monitor and a different volume measured on the BL first APA. Full uncertainties are shown.}
    \label{fig:allThree}
\end{figure}

\begin{figure}
    \centering
    \includegraphics[width=0.65\textwidth]{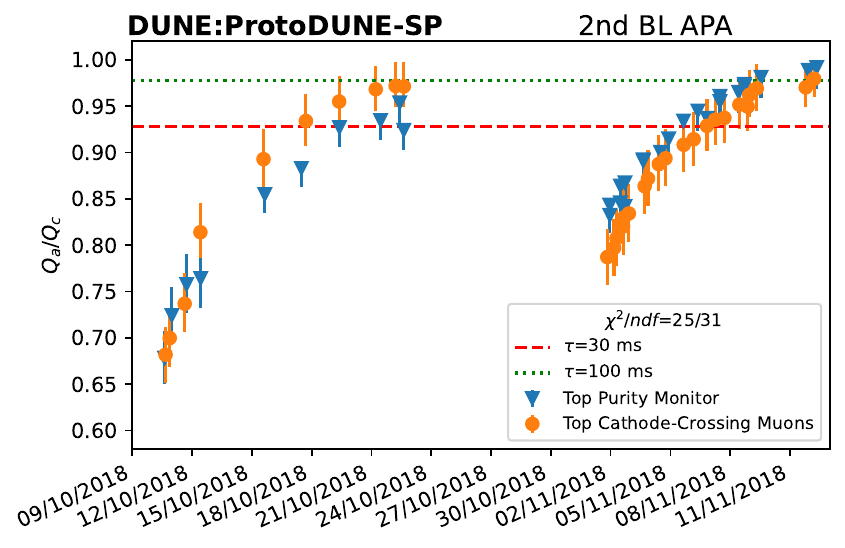}
        \includegraphics[width=0.65\textwidth]{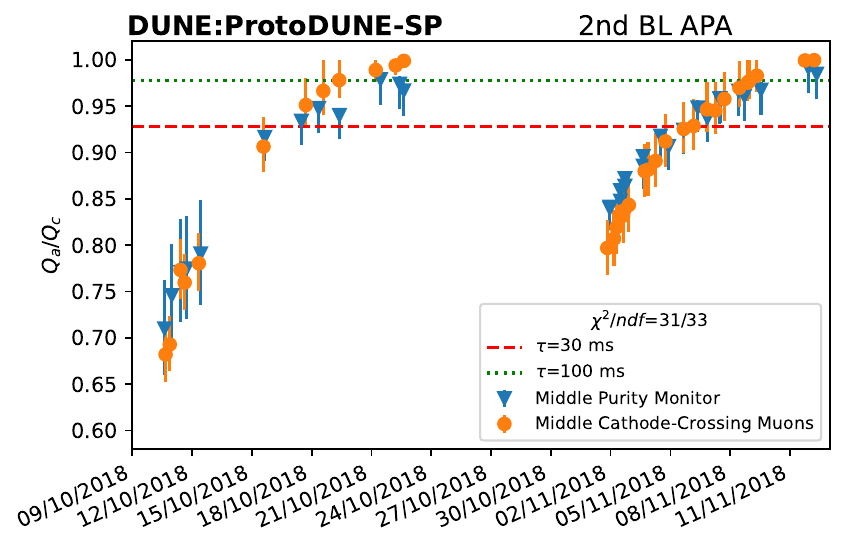}
    \includegraphics[width=0.65\textwidth]{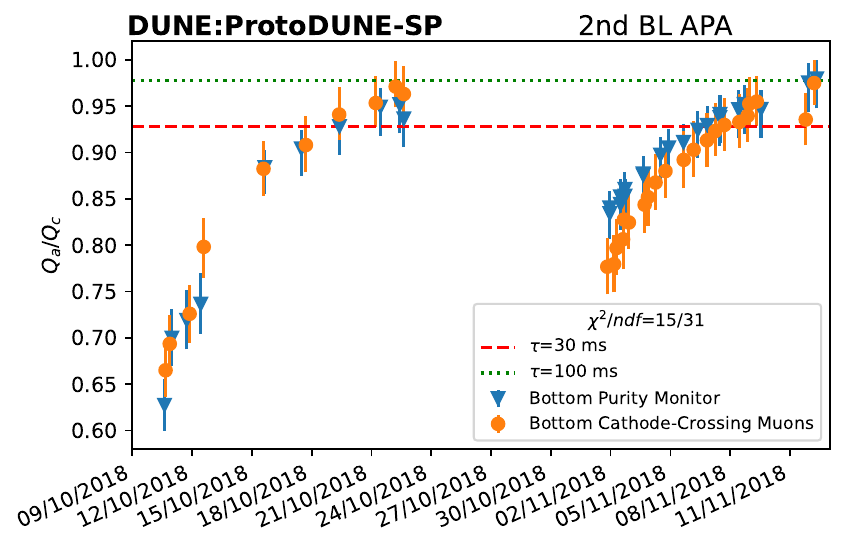}
    \caption{Measurements of the minimum fraction of charge remaining from cathode-crossing muons in the second APA and the purity monitors. Full uncertainties are shown.}
    \label{fig:allThree2}
\end{figure}

\begin{figure}
    \centering
    \includegraphics[width=0.65\textwidth]{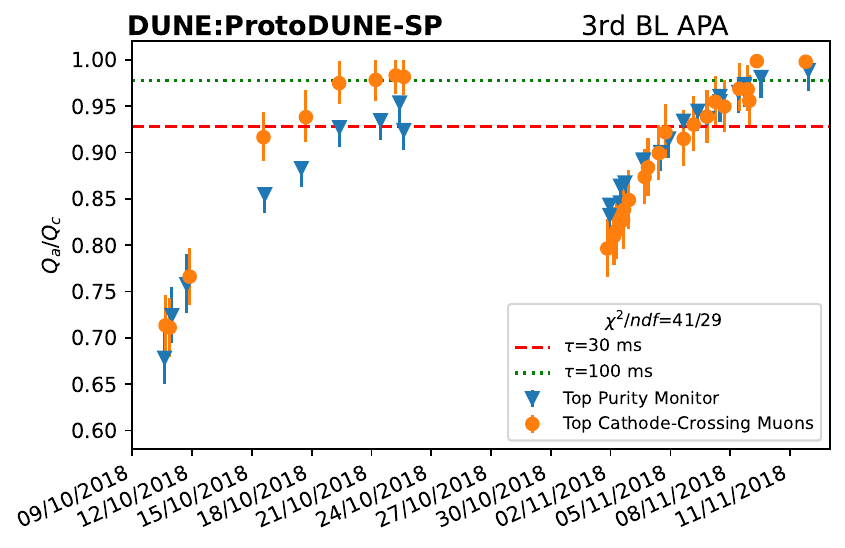}
        \includegraphics[width=0.65\textwidth]{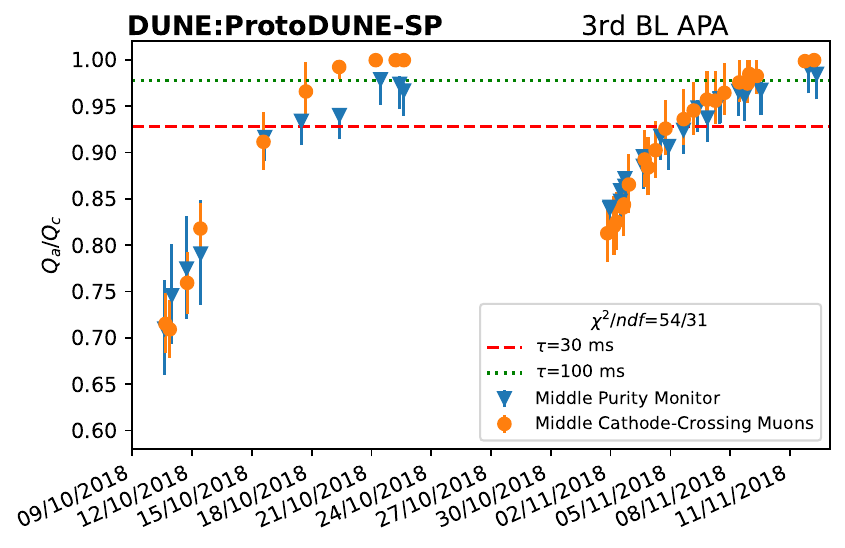}
    \includegraphics[width=0.65\textwidth]{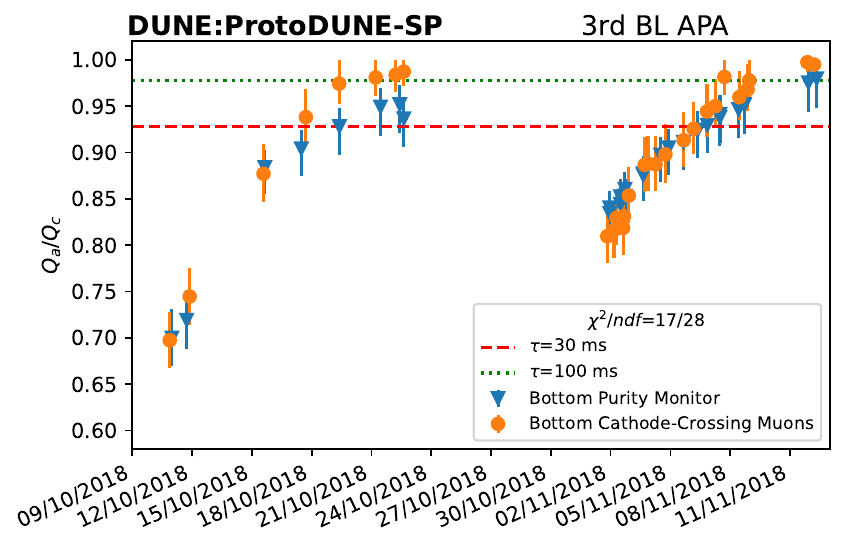}
    \caption{Measurements of the minimum fraction of charge remaining from cathode-crossing muons in the third APA and the purity monitors. Full uncertainties are shown.}
    \label{fig:allThree3}
\end{figure}

To compare to the TPC data, the purity monitor data requires adjustments to account for the higher electric field. Measurements of $\rm{\frac{Q_a}{Q_c}}$ from the purity monitor are scaled to the electric field of the TPC by recalculating the attachment rate using~\cite{thorn}, assuming that the impurities are due to oxygen based on investigations of the recirculation and filtration system. The correction, which uses collected data of argon purity as a function of impurity concentration, ensures the drift electron lifetime measurements from the purity monitors can be applied to calibrate the TPC. The purity monitors are installed outside the field cage in the upstream BL corner. While there are low statistics in that region, as it is a small corner of the detector, evaluations were made of the relevant volume near the purity monitors, the first APA on the BL side, using the cathode-crossing event selection. Measurements were made at each height over roughly one month of consistent operation. Figure~\ref{fig:allThree} shows the results. The first APA is a challenging area to measure the drift electron lifetime due to the SCE that distorts signals on the front face, forcing the analysis to have a reduced volume for signals in the first APA and, therefore, less statistics as outlined in Table~\ref{tab:volumeZ}. Comparisons were also made with the other APAs. Figure~\ref{fig:allThree2} and Figure~\ref{fig:allThree3} use the cathode-crossing selection to measure the charge remaining in the second and third BL APAs, showing similar trends in argon purity over time. 
 
 As a reminder, data points in the chi-square statistic must be within six hours of each other to compare at similar periods of operation. The highest disagreements are at the top and bottom areas of the volume in the first APA, where the purity monitors sit, and are driven by the lack of agreement at the end of October, when the TPC measured near infinite lifetimes. However, the ``central-value'' measurements, where the SCE is minimal and statistics are larger in the second APA for the cathode-crossing selection, have good agreement with the purity monitor data with a chi-square per degrees of freedom statistic of approximately 1. Regardless, all measurements maintained high drift electron lifetimes above DUNE technical specifications of 10 ms, with it often times greater than 30 ms and occasionally reaching 100 ms. 

Given the low cosmic-ray muon flux expected underground, the purity monitors are the primary devices to determine the liquid argon purity for the DUNE Far Detector modules. This study shows that the purity monitors and the TPC both report similar levels of argon purities. In October 2018, the TPC reported much higher purities than the purity monitor. Although some measurements in November 2018 reported lower purities from the TPC, those specific comparisons still agreed within an approximate 1$\sigma$-level according to the chi-squared values calculated. Any statistically significant disagreements occurred when the TPC reported significantly higher lifetimes than the purity monitors, as seen when comparing to the bottom purity monitor and data from the first BL APA in Figure~\ref{fig:allThree}, implying the purity monitors can provide conservative evaluations of the drift electron lifetime without overstating performance. Finally, the TPC shows the same general trend of the purity monitors with most measurements above the 10 ms technical specification ($Q_a/Q_c$=0.80) and all measurements above the 3 ms technical requirement ($Q_a/Q_c$=0.47).

\section{Conclusion} \label{sec:conclusion}

The paper summarizes and compares the analyses of the liquid argon purity using the TPC and the purity monitors of ProtoDUNE-SP. The DUNE Far Detector modules aim to achieve a drift electron lifetime of over 10 ms. Over a steady operating period from October to November 2018, ProtoDUNE-SP measured high drift electron lifetimes with a percent of charge recovered consistently over 95\%, which translates to a drift electron lifetime of over 50 ms, and a lifetime reaching above 100 ms. The data proves the feasibility of achieving a 10 ms drift electron lifetime given similar engineering and designs of ProtoDUNE-SP. Due to the low cosmic-ray flux at the DUNE Far Detector, the purity monitor data will be the dominant method for calibrating impurities and this paper shows good agreement between the purity monitor data and TPC data. These results illustrate that the purity of the liquid argon can be measured with sufficient accuracy to calibrate the charge loss due to the capture of the ionization electrons by impurities. Using Equation~\ref{eqn:qfracuncert}, the uncertainty on the charge measured from argon impurity calibrations is at the sub-percent level from TPC data under stable operating conditions.


The analyses developed for this publication will form the basis of future analyses for the ProtoDUNE Horizontal Drift and ProtoDUNE Vertical Drift modules. They will allow for quick evaluations of the liquid argon purity in the TPC and probe the liquid argon purity for the much larger drift distance of the DUNE Vertical Drift module, which has a drift length of over 6 m~\cite{vdTDR}. Improvements can be made for TPC measurements by constraining the drift electron diffusion systematic uncertainty, a major unknown in liquid argon detector physics, and increasing the run period of steady-state operation of the TPC, which for ProtoDUNE-SP only lasted for one month in 2018.

\section*{Acknowledgements}

The ProtoDUNE-SP detector was constructed and operated on the CERN Neutrino Platform.
We gratefully acknowledge the support of the CERN management, and the
CERN EP, BE, TE, EN and IT Departments for NP04/Proto\-DUNE-SP.
%
%
This document was prepared by DUNE collaboration using the resources of the Fermi National Accelerator Laboratory (Fermilab), a U.S. Department of Energy, Office of Science, Office of High Energy Physics HEP User Facility. Fermilab is managed by Fermi Forward Discovery Group, LLC, acting under Contract No. 89243024CSC000002.
%
%
This work was supported by
CNPq,
FAPERJ,
FAPEG and 
FAPESP,                         Brazil;
CFI, 
IPP and 
NSERC,                          Canada;
CERN;
M\v{S}MT,                       Czech Republic;
ERDF, FSE+,
Horizon Europe, 
MSCA and NextGenerationEU,      European Union;
CNRS/IN2P3 and
CEA,                            France;
PRISMA+,                        Germany;
INFN,                           Italy;
FCT,                            Portugal;
NRF,                            South Korea;
Generalitat Valenciana, 
Junta de Andaluc\'ia-FEDER, 
MICINN, and 
Xunta de Galicia,               Spain;
SERI and 
SNSF,                           Switzerland;
T\"UB\.ITAK,                    Turkey;
The Royal Society and 
UKRI/STFC,                      United Kingdom;
DOE and 
NSF,                            United States of America.

\bibliography{sample.bib}
\bibliographystyle{JHEP.bst}

\end{document}